\documentclass[aps,pre,reprint,showpacs,superscriptaddress,groupaddress,showkeys,floatfix,byrevtex]{revtex4-1}
\usepackage{amsmath}
\usepackage{amsfonts}
\usepackage{amssymb}
\usepackage[dvipsnames]{xcolor}
\usepackage{graphicx,color,subeqnarray}
\usepackage{dcolumn}
\usepackage{bm}
\usepackage{srcltx}
\usepackage{soul}

\begin{document}

\title{Two-dimensional active motion}

\author{Francisco J. Sevilla}
\email{fjsevilla@fisica.unam.mx}
\thanks{corresponding author}
\affiliation{Instituto de F\'isica, Universidad Nacional Aut\'onoma de M\'exico,
Apdo.\ Postal 20-364, 01000, Ciudad de M\'exico, M\'exico
}
\keywords{\href{https://doi.org/10.1103/PhysRevE.101.022608}{DOI: 10.1103/PhysRevE.101.022608}}

\date{\today}

\begin{abstract}
The diffusion in two dimensions of noninteracting active particles that follow an arbitrary motility pattern is considered for analysis. {A Fokker-Planck-like} 
equation is generalized to take into account an arbitrary distribution of scattered angles of the swimming direction, which encompasses the pattern of {active} motion of 
particles that move at constant speed. An exact analytical expression for the marginal probability density of finding a particle on a given position at a given instant, 
independently of its direction of motion is provided, and a connection with a generalized diffusion equation is unveiled. Exact analytical expressions for the time dependence 
of the mean-square displacement and of the kurtosis of the distribution of the particle positions are presented. {The analysis is focused in the intermediate-time regime, 
where the effects of the specific pattern of active motion are conspicuous.} For this, it is shown that only the {expectation value of the} first two {harmonics of the 
scattering angle of the} direction of motion are needed. The effects of persistence and of circular motion are discussed for different families of distributions of 
the scattered direction of motion. 
\end{abstract}

\maketitle

\section{Introduction}
The intens{ive} study of the out-of-equilibrium systems called \emph{active matter}, has allowed to set up a firm basis for the understanding  of a variety of 
out-of-equilibrium phenomena. Even at the individual level of description, the intrinsic nonequilibrium nature of active motion leads to diverse  
phenomena not observed in particles that move passively. Furthermore, the great diversity of the patterns of self-propelled motion 
observed in biological organisms (see the introductory section in Refs. \cite{TaktikosPlos2014,DetcheverryEPL2015}) or in artificially designed active particles 
\cite{BechingerRMP2016}, enriches the variety of effects exhibited by these systems . 

A salient feature of active motion is that it is \emph{persistent}, a characteristic that explicitly depends on the specific pattern of motion performed by the 
particle.  The effects of persistence are well known, for instance, when active particles are confined to move under the effects of either an external potential or  
hard-walls, to lead to stationary distributions that differ from the ones {of passive particles}. In the case of confined motion by trapping potentials, 
the effects of persistence lead to distributions that differ from the one given by Boltzmann and Gibbs 
{\cite{PototskyEPL2012,CatesEPL2013,Solon2015EPJST2015,StarkEPJST2016,CapriniSciRep2019}}. 

On the other hand, {among the many models that describe active motion \cite{WuPRL2000,SzamelPRE2014,FodorPRL2016,MaggiSciRep2015,FaragePRE2015,PaoluzziPRE2016,Sevilla2019c},} 
theoretical comparative studies that consider {the free diffusion of} two {of the} more studied patterns of active motion --\emph{active 
Brownian motion}, where the orientation of motion undergoes rotational diffusion, and \emph{run-and-tumble motion}, which alternates running events with instantaneous, 
temporally uncorrelated tumbling events-- reveal that, although there are important quantitative differences between them {in the 
intermediate-time regime}, they behave similarly in the long-time regime, {namely, the exhibit normal diffusion}, and have the same behavior in the short-time 
regime, {that is, they move ballistically}. 

It is precisely in the intermediate-time regime, i.e., for times of the order of the \emph{persistence time}, when conspicuous differences are revealed between both patterns 
of motion \cite{KurzthalerPRL2018}. {Based on these findings, I present in this paper an analysis of the statistics of active 
motion of particles that follows an arbitrary, however Markovian and spatially local, orientational dynamics. I focus the analysis to the intermediate-time regime, where the 
difference among specific patters of motion are conspicuous, as shown in the following sections, naturally, Gaussian normal diffusion is observed in the long-time regime and 
non-Gaussian ballistic superdiffusion in the short-time one.}

Hence, to have at our disposal a theoretical framework that incorporates an arbitrary pattern of motion {(circular, run-and-reverse, run-and-flick, etc.)} of 
active swimmers it is highly desirable. In this paper, I present a theoretical framework of two-dimensional motion of active swimmers for a family of patterns of motion 
characterized by constant speed and an arbitrary probability distribution of the turning angle  (scattered angle) of the swimming direction. Another important theoretical 
framework based on the \emph{continuous time random walks} of single particles has been known in the literature, {however} this focuses on a family of 
patterns of motion characterized by the Poissonian or non-Poissonian statistics between turning events \cite{TaktikosPlos2014,DetcheverryEPL2015,DetcheverryPRE2017}. 

On the basis of the \emph{transport equation} \cite{DuderstadtTransportTheory,PorraPRE1997}, I introduce in section \ref{SectModel} such a framework, and present the 
corresponding Fokker-Planck equation for the probability density that at time $t$, a particle is located at $\boldsymbol{x}$ and moving along the direction 
$\hat{\boldsymbol{v}}$. {Although this theoretical framework considers the case of a spatially local Markovian dynamics of a single active particle, this 
is susceptible to be generalized not only to incorporate a more complex dyamics (as suggested by recent experiments on \emph{Escherichia Coli} in Ref. 
\cite{Figueroa2018}), like memory effects in the swimming dynamics \cite{SevillaPRE2019b},} {but also, to 
include many-body interactions, which in combination with a specific pattern of active motion, have important consequences in the collective dynamics 
\cite{ZhangPNAS2010,ThutupalliJRSI2015,GinotNatureComm2018,MahaultPRL2018,JeckelPNAS2018,LiuPRL2019}.}
The general solution of such a Fokker-Planck equation is presented in Sec. \ref{SectGralSolution}. The marginal probability distribution of finding a swimmer at 
$\boldsymbol{x}$ at time $t$, independently 
of the direction of motion is of great interest, and in Sec. \ref{SectGralSolution} I provide an exact solution, whose physical consequences are analyzed. A 
connection with a \emph{generalized diffusion equation} is also unveiled. In Sec \ref{SectParticularCases}, generalities, applications and predictions of the framework 
are presented for some general families of patterns of motion. Finally I give my concluding remarks in Sec. \ref{SectConclusion}.

\section{\label{SectModel}The two-dimensional active transport equation}

The starting point is the two-dimensional equation for the probability
density, $\mathcal{P}({\boldsymbol{x}},\varphi ,t)$, of a single particle being at position $\boldsymbol{x}$, moving at constant speed $v_{0}$ along a direction given by 
the angle $\varphi$ at time $t$, that is,
\begin{multline}\label{TransportEquation}
\frac{\partial }{\partial t}\mathcal{P}({\boldsymbol{x}},\varphi ,t)+v_{0}\hat{\boldsymbol{v}}\cdot \nabla \mathcal{P}({\boldsymbol{x}},\varphi 
,t)=D_{T}\nabla^{2}\mathcal{P}({\boldsymbol{x}},\varphi ,t)\\
+\int_{-\pi}^{\pi}d\varphi^{\prime} K_A\left(\varphi\vert\varphi^{\prime}\right)\mathcal{P}(\boldsymbol{x},\varphi^{\prime},t),
\end{multline}
where the unit vector $\hat{\boldsymbol{v}}$ is defined by $(\cos\varphi,\sin\varphi)$, $\varphi$ being the angle between the {particle} direction of motion and the 
horizontal axis of a given Cartesian reference frame. $D_{T}$ is the translational diffusion coefficient that gives account of the thermal fluctuations exerted by the 
surrounding medium.  The transition rate of the direction of motion, $K_{A}(\varphi\vert\varphi^{\prime})$, gives the probability rate of the transition from the direction of 
motion $\varphi^{\prime}$ to $\varphi$, and encompasses the detailed information of a specific pattern of active motion considered. In this paper, I focus on the {broad} 
case in which $K_{A}(\varphi\vert\varphi^{\prime})$ is independent of time, of the swimming speed and of the particle position. However, 
such dependences must be considered in the more general {case}, as for instance {to describe the motion} of the bacterium 
\emph{Pseudomonas putida}, whose swimming speed depends on the selected direction of motion after a transition \cite{ThevesBiophysJ2013}{, or the motion of 
\emph{E. coli}, for which a large variability in its motility behavior has been observed \cite{Figueroa2018}.} 

I refer to Eq. \eqref{TransportEquation} as the \emph{active-transport equation}. The {fact that} passive fluctuations exerted on the particle motion are separated 
from the active ones allows to write {(see the Appendix)}
\begin{equation}\label{SolutionSeparation}
 \mathcal{P}(\boldsymbol{x},\varphi,t)=\int d^{2}x^{\prime}\, G_{D_{T}}(\boldsymbol{x}-\boldsymbol{x}^{\prime},t)P(\boldsymbol{x}^{\prime},\varphi,t),
\end{equation}
where $G_{D_{{T}}}(\boldsymbol{x},t)$ denotes the two-dimensional Gaussian propagator of the diffusion equation with diffusion coefficient 
$D_{{T}}$, given explicitly by $\exp\{-\boldsymbol{x}^{2}/4D_{{T}}t\}/4\pi D_{{T}}t$. The active part of motion is entailed by the probability 
density 
$P(\boldsymbol{x},\varphi,t)$,  which 
satisfies 
the gain-loss equation
\begin{multline}\label{TransportEquation2}
\frac{\partial }{\partial t}P({\boldsymbol{x}},\varphi ,t)+v_{0}\hat{\boldsymbol{v}}\cdot \nabla P({\boldsymbol{x}},\varphi ,t)=\\
\int_{-\pi}^{\pi}d\varphi^{\prime} Q\left(\varphi,\varphi^{\prime}\right)P(\boldsymbol{x},\varphi^{\prime},t)\\
-\left[\int_{-\pi}^{\pi}d\varphi^{\prime} Q\left(\varphi^{\prime},\varphi\right)\right]P(\boldsymbol{x},\varphi,t),
\end{multline}
when $K_{A}(\varphi\vert\varphi^{\prime})$ is written in terms of the distribution of scattering-angle $Q(\varphi,\varphi^{\prime})$ as
\begin{equation}\label{GainLossTransitions}
 K_{A}(\varphi\vert\varphi^{\prime})=Q(\varphi,\varphi^{\prime})-\delta(\varphi-\varphi^{\prime})\int_{-\pi}^{\pi}d\varphi^{\prime\prime}Q(\varphi^{\prime\prime},\varphi).
\end{equation}

A further simplification can be realized by considering a rotationally invariant transition rate function, i.e., $Q(\varphi,\varphi^{\prime})=Q(\varphi-\varphi^{\prime})$. In 
such a case we can write \cite{PorraPRE1997}
\begin{multline}\label{TransportEquation3}
\frac{\partial }{\partial t}P({\boldsymbol{x}},\varphi ,t)+v_{0}\hat{\boldsymbol{v}}\cdot \nabla P({\boldsymbol{x}},\varphi ,t)=\\
\Lambda\int_{-\pi}^{\pi}d\varphi^{\prime} \widetilde{Q}\left(\varphi-\varphi^{\prime}\right)P(\boldsymbol{x},\varphi^{\prime},t)\\
-\Lambda P(\boldsymbol{x},\varphi,t),
\end{multline}
where $\Lambda\equiv\int_{-\pi}^{\pi}d\varphi^{\prime}Q(\varphi^{\prime})$ is the inverse of the timescale that measures the average time between transitions, and 
$\widetilde{Q}(\varphi)=Q(\varphi)/\Lambda$.  

\section{\label{SectGralSolution}The general solution to the active transport equation}
We are interested in the analytical solutions, $P(\boldsymbol{x},\varphi,t)$, if any, of Eq.~(\ref{TransportEquation3}), 
with the initial condition $P({\boldsymbol{x}},\varphi ,0)=\delta ^{(2)}({\boldsymbol{x}})/2\pi $, which corresponds to the case of an ensemble of independent active 
particles that depart from the origin in a Cartesian system of coordinates, and propagates in a random direction of motion drawn from the uniform distribution in 
$[-\pi,\pi]$, 
$\delta ^{(2)}({\boldsymbol{x}})$ being the two dimensional Dirac's delta function. 

Due to the assumed spatial isotropy of the system, I apply the Fourier transform to Eq. (\ref{TransportEquation3}) and obtain 
\begin{multline}\label{Transformed}
\frac{\partial }{\partial t}\widetilde{P}({\boldsymbol{k}},\varphi ,t)+iv_{0}\,\hat{\boldsymbol{v}}\cdot {\boldsymbol{k}}\, \widetilde{P}({%
\boldsymbol{k}},\varphi ,t) =\\
\Lambda\int_{-\pi}^{\pi}d\varphi^{\prime} \widetilde{Q}\left(\varphi-\varphi^{\prime}\right)\widetilde{P}(\boldsymbol{k},\varphi^{\prime},t)\\
-\Lambda \widetilde{P}(\boldsymbol{k},\varphi,t),
\end{multline}
where
\begin{equation}
\widetilde{P}({\boldsymbol{k}},\varphi ,t)=\int \frac{d^{2}x}{2\pi}\,e^{-i\boldsymbol{k}\cdot \boldsymbol{x}}\,P(\boldsymbol{x},\varphi ,t),
\label{fourier}
\end{equation}
denotes the symmetric Fourier transform of $P(\boldsymbol{x},\varphi,t)$ and ${\boldsymbol{k}}=(k_{x},k_{y})$, denotes the system's wave-vector. The following Fourier 
series expansion,
\begin{equation}\label{Pexpansion}
\widetilde{P}({\boldsymbol{k}},\varphi ,t)=\frac{1}{2\pi}\sum\limits_{n=-\infty}
^{\infty }\widetilde{p}_{n}({\boldsymbol{k}},t)\, e^{-\lambda_{n}t}\, e^{in\varphi},  
\end{equation}
is suitable since it fulfills the periodicity condition of the probability density, $\widetilde{P}({\boldsymbol{k}},\varphi 
,t)=\widetilde{P}({\boldsymbol{k}},\varphi+2\pi,t)$.

The coefficients $\widetilde{p}_{n}(\boldsymbol{k},t)$ in the expansion \eqref{Pexpansion} are obtained by the use of the 
standard orthogonality relation among the Fourier basis functions $\left\{e^{in\varphi}\right\}$, explicitly
\begin{align}
\widetilde{p}_{n}({\boldsymbol{k}},t)&=\int \frac{d^{2}x}{2\pi}\, e^{-i\boldsymbol{k}\cdot\boldsymbol{x}}\, p_{n}(\boldsymbol{x},t)\\
&=e^{\lambda_{n}t}\int_{-\pi}^{\pi}d\varphi
\, \widetilde{P}({\boldsymbol{k}},\varphi ,t)e^{-in\varphi }  \label{inv}
\end{align}
and satisfy the identity $\widetilde{p}_{-n}^{*}(\boldsymbol{k},t)=\widetilde{p}_{n}(\boldsymbol{k},t)$, since the probability density $P(\boldsymbol{x},\varphi,t)$ is 
a real function. The factors $e^{-\lambda_{n}t}$ in the expansion \eqref{Pexpansion}, correspond to the coefficients, $c_{n}(t)$, of the expansion in Fourier series of 
$f(\varphi,t)$ 
that solves the equation
\begin{multline}
\frac{\partial }{\partial t}f(\varphi,t)=\Lambda\int_{-\pi}^{\pi}d\varphi^{\prime} \widetilde{Q}\left(\varphi-\varphi^{\prime}\right)f(\varphi^{\prime},t)
-\Lambda f(\varphi,t),
\end{multline}
with $\lambda_{n}$ a complex number given by
\begin{equation}
 \lambda_{n}=\Lambda\left[1-\langle e^{-in\varphi}\rangle_{\widetilde{Q}}\right],
\end{equation}
where
\begin{equation}
 \langle\Phi(\varphi)\rangle_{\widetilde{Q}}=\int_{-\pi}^{\pi}d\varphi\, \widetilde{Q}(\varphi) \Phi(\varphi)
\end{equation}
denotes the average of the $\varphi$-dependent quantity $\Phi(\varphi)$ computed by the use of the scattering-angle distribution $\widetilde{Q}(\varphi)$.

Accordingly, the main features of a particular pattern of active motion are encoded in the distribution of scattered angles  $\widetilde{Q}(\varphi)$, which 
{entails} the 
particular orientation {dynamics} of the swimming direction. Such features are equivalently 
inherited in the trigonometric moments:
\begin{subequations}\label{lambda-n}
 \begin{align}
  \Gamma_{n}&=\Lambda\left[1-\langle\cos n\varphi\rangle_{\widetilde{Q}}\right],\label{Gamma-n}\\
  \Omega_{n}&=\Lambda \langle\sin n\varphi\rangle_{\widetilde{Q}},\label{Omega-n}
  \end{align}
\end{subequations}
which correspond to the real and imaginary part of $\lambda_{n}$, respectively, thus $\lambda_{n}=\Gamma_{n}+i\, \Omega_{n}$. These quantities 
{fully characterize} { the statistical properties of active motion} (see {for 
instance} Refs. \cite{ViswanathanPRE2005,BartumeusJTB2008} {where only $\langle\cos\varphi\rangle_{\widetilde{Q}}$ is considered for their analysis 
of two-dimensional} \emph{correlated random walks}). 

A series of properties for $\Gamma_{n}$ and $\Omega_{n}$ can be deduced in a straightforward way. From the 
normalization of $\widetilde{Q}$ we have that $\Gamma_{0}=\Omega_{0}=0$, and since $\widetilde{Q}(\varphi)$ is a real-valued function, we have that the complex conjugate of 
$\lambda_{n}$ is given by $\lambda_{n}^{*}=\lambda_{-n}$, which implies $\Gamma_{n}=\Gamma_{-n}$ and $\Omega_{n}=-\Omega_{-n}$. From this property one can show that the 
coefficients $\widetilde{p}_{n}(\boldsymbol{k},t)$ of the expansion 
\eqref{Pexpansion2} satisfy $\widetilde{p}_{-n}(\boldsymbol{-k},t)=\widetilde{p}^{*}_{n}(\boldsymbol{k},t)$. Notice further that $0\le\Gamma_{n}\le2\Lambda$ 
and that $-\Lambda\le\Omega_{n}\le\Lambda$. With this observations, the expansion 
\eqref{Pexpansion} can be explicitly split as 
\begin{multline}\label{Pexpansion2} 
\widetilde{P}(\boldsymbol{k},\varphi,t)=\frac{1}{2\pi}\widetilde{p}_{0}(\boldsymbol{k},t)+\\ 
\frac{1}{2\pi}\sum_{\substack{n=-\infty,\\n\neq 0}}^{\infty}\widetilde{p}_{n}(\boldsymbol{k},t)e^{ -\Gamma_ { n } t}e^{ -i\Omega_ { n }
 t}e^{in\varphi},
\end{multline}
where {$\Gamma_{n}$ expresses the damping rate of the contribution of the $n$-th Fourier mode in the expansion \eqref{Pexpansion}.} 
$P(\boldsymbol{x},\varphi,t)$ tends asymptotically to $p_{0}(\boldsymbol{x},t)/2\pi$ as time goes by{, since $\Gamma_{0}=0$ and $\Omega_{0}=0$.}

\subsection{\label{SubSectIIIA}The coefficients $p_{n}(\boldsymbol{x},t)$}
After substitution of Eq.~\eqref{Pexpansion} into Eq.~\eqref{Transformed}, and use of the orthogonality of the Fourier basis functions, a set of coupled 
ordinary differential equations for the coefficients $\widetilde{p}_{n}({
\boldsymbol{k}},t)$ is obtained, namely \cite{SevillaPRE2014, SevillaPRE2015, SevillaPRE2016}
\begin{multline}\label{Hierarchy}
\frac{d}{dt}\widetilde{p}_{n}(\boldsymbol{k},t) =-\frac{v_{0}}{2}ike^{\lambda_{n}t}\left[
e^{-i\theta}\,e^{-\lambda_{n-1}t}\, \widetilde{p}_{n-1}(\boldsymbol{k},t)\right.\\
\left.+e^{i\theta}\,e^{-\lambda_{n+1}t}\, \widetilde{p}_{n+1}(\boldsymbol{k},t)\right],
\end{multline}
where $\theta$ and $k$ correspond to the polar coordinates of the two-dimensional Fourier vector $\boldsymbol{k}$, i.e.,  $k_{x}\pm ik_{y}=ke^{\pm i\theta }$. Equations 
\eqref{Hierarchy} are complemented by the initial conditions $\widetilde{p}_{n}^{(0)}(\boldsymbol{k})=(2\pi)^{-1}\delta_{n,0}$, which are obtained straightforwardly 
from the initial distribution considered, i.e., $P(\boldsymbol{x},\varphi,0)=\delta^{(2)}(\boldsymbol{x})/2\pi$.

{Notice that} {t}he first coefficient $p_{0}(\boldsymbol{x},t)$ {is} related to the probability density 
\begin{equation}
 \varrho(\boldsymbol{x},t)=\frac{1}{2\pi}p_{0}(\boldsymbol{x},t)=\frac{1}{2\pi}\int_{-\pi}^{\pi}d\varphi\, P(\boldsymbol{x},\varphi,t).
\end{equation}
{The next coefficients, $p_{\pm1}(\boldsymbol{x},t)$,} {are related to} the first-rank tensor $\boldsymbol{j}(\boldsymbol{x},t)$ {whose} 
components {are given by}
\begin{subequations}
\begin{align}
 j_{x}(\boldsymbol{x},t)&=\frac{{e^{-\Gamma_{1}t}}}{\pi}\text{Re}[p_{1}(\boldsymbol{x},t){e^{-i\Omega_{1}t}}] \nonumber\\
 &=\frac{1}{\pi}\int_{-\pi}^{\pi}d\varphi\, \cos\varphi\, P(\boldsymbol{x},\varphi,t),\\ 
 j_{y}(\boldsymbol{x},t)&=\frac{{e^{-\Gamma_{1}t}}}{\pi}\text{Im}\left[p_{-1}(\boldsymbol{x},t){e^{i\Omega_{1}t}}\right]\nonumber\\
 &=\frac{1}{\pi}\int_{-\pi}^{\pi}d\varphi\, \sin\varphi\, P(\boldsymbol{x},\varphi,t).
\end{align}
\end{subequations}
{These, give the average direction of motion at position $\boldsymbol{x}$ at time $t$ and} from which the probability 
density current $\boldsymbol{J}(\boldsymbol{x},t)=\frac{v_{0}}{2}\boldsymbol{j}(\boldsymbol{x},t)$ is introduced.  {$\text{Re[z]}$ and $\text{Im}[z]$ denote 
for the real and imaginary part of $z$ respectively.}

{The coefficients $p_{\pm2}(\boldsymbol{x},t)$} {define} the traceless, symmetric, 2$\times$2 second-rank tensor $\mathbb{W}(\boldsymbol{x},t)$, 
whose entries are given by
\begin{subequations}
\begin{align}
 \mathbb{W}_{xx}(\boldsymbol{x},t)&=-\mathbb{W}_{yy}(\boldsymbol{x},t)\nonumber\\
 &=\frac{{e^{-\Gamma_{2}t}}}{\pi}\text{Re}\left[p_{2}(\boldsymbol{x},t){e^{-i\Omega_{2}t}}\right]\nonumber\\
 &=\frac{1}{\pi}\int_{-\pi}^{\pi}d\varphi\, \cos2\varphi\, P(\boldsymbol{x},\varphi,t),\\
 \mathbb{W}_{xy}(\boldsymbol{x},t)&=\mathbb{W}_{yx}(\boldsymbol{x},t)\nonumber\\
 &=\frac{{e^{-\Gamma_{2}t}}}{\pi}\text{Im}\left[p_{-2}(\boldsymbol{x},t){e^{i\Omega_{2}t}}\right]\nonumber\\
 &=\frac{1}{\pi}\int_{-\pi}^{\pi}d\varphi\, \sin2\varphi\, P(\boldsymbol{x},\varphi,t).
\end{align}
\end{subequations}
{The matrix
\begin{align}
\mathbb{R}(\boldsymbol{x},t)=&\frac{\mathbb{W}(\boldsymbol{x},t)}{\mathbb{W}_{xx}^{2}(\boldsymbol{x},t)+\mathbb{W}_{xy}^{2}(\boldsymbol{x},t)}\nonumber\\
=&\left(\begin{array}{cc}
  \cos2\Theta(\boldsymbol{x},t) & \sin2\Theta(\boldsymbol{x},t)\\
  \sin2\Theta(\boldsymbol{x},t)  & -\cos2\Theta(\boldsymbol{x},t)
  \end{array}\right),
\end{align}
corresponds to the two-dimensional matrix representation of the \emph{reflection transformation} about the direction 
$\hat{\boldsymbol{r}}(\boldsymbol{x},t)=\bigl(\cos\Theta(\boldsymbol{x},t),\sin\Theta(\boldsymbol{x},t)\bigr)$, where $\Theta(\boldsymbol{x},t)$ is given by}
{
\begin{equation}
\tan 2\Theta(\boldsymbol{x},t)=\frac{\mathbb{W}_{xy}(\boldsymbol{x},t)}{\mathbb{W}_{xx}(\boldsymbol{x},t)}.
\end{equation}
}

{With $\varrho(\boldsymbol{x},t)$, $\boldsymbol{j}(\boldsymbol{x},t)$, $\mathbb{W}(\boldsymbol{x},t)$ and so on,  it is customarily to rewrite 
$\widetilde{P}(\boldsymbol{k},\varphi,t)$ in the form}
{
\begin{equation}\label{Pexpansion3}
\widetilde{P}({\boldsymbol{k}},\hat{\boldsymbol{v}} 
,t)=\widetilde{\varrho}(\boldsymbol{k},t)+\hat{\boldsymbol{v}}\cdot\widetilde{\boldsymbol{j}}(\boldsymbol{k},t)+\hat{\boldsymbol{v}}\cdot\widetilde{\mathbb{W}}
(\boldsymbol { k } , t)\cdot\hat{\boldsymbol{v}}+\ldots,
\end{equation}
}{where the second term in the right-hand side gives the contribution to $P(\boldsymbol{x},\hat{\boldsymbol{v}},t)$ due to the projection of this average direction of motion along the direction of 
motion $\hat{\boldsymbol{v}}$. This term decays exponentially at the rate $\Gamma_{1}$, whose inverse characterizes the \emph{persistence time} of active motion. The third 
term in the right-hand side of 
\eqref{Pexpansion3}, gives a contribution to $P(\boldsymbol{x},\hat{\boldsymbol{v}},t)$ proportional to the projection of the reflected direction of motion 
$\mathbb{R}(\boldsymbol{x},t)\hat{\boldsymbol{v}}$ [about the axis $\hat{\boldsymbol{r}}(\boldsymbol{x},t)$], along $\hat{\boldsymbol{v}}$. This term decays exponentially at 
the rate $\Gamma_{2}$. As is shown afterwards in the following sections, $\Gamma_{1}^{-1}$, $\Gamma_{2}^{-1}$, $\Omega_{1}$ and $\Omega_{2}$, 
are necessary to give a minimal comprehensive description of the statistical properties of active motion.}  

\subsection{The probability density $p_{0}(\boldsymbol{x},t)$}
As in previous studies, the probability density of finding a particle at position $\boldsymbol{x}$, independently of its direction of motion, $p_{0}(\boldsymbol{x},t)$, is of 
interest. After transforming the time domain to the Laplace domain, an exact solution for 
{
\begin{equation}\label{p0FourierLaplace}
\widetilde{p}_{0}(\boldsymbol{k},\epsilon)=\int\frac{d^{2}x}{2\pi}e^{-i\boldsymbol{k}\cdot\boldsymbol{x}}p_{0}(\boldsymbol{k},\epsilon) 
\end{equation}
}can be obtained from 
Eq. \eqref{Hierarchy} in the form of continuous fractions {(see the Appendix \ref{Appendix})}, namely
\begin{widetext}
\begin{equation}\label{p0}
\widetilde{p}_{0}(\boldsymbol{k},\epsilon)=\widetilde{p}_{0}^{(0)}(\boldsymbol{k})\cfrac{1}{\epsilon
	  + \cfrac{(v_{0}/2)^{2}\boldsymbol{k}^{2}}{\epsilon+\lambda_{1}
          + \cfrac{(v_{0}/2)^{2}\boldsymbol{k}^{2}}{\epsilon+\lambda_{2}
          + \cfrac{(v_{0}/2)^{2}\boldsymbol{k}^{2}}{\epsilon+\lambda_{3}+\ddots
           } } }+
	    \cfrac{(v_{0}/2)^{2}\boldsymbol{k}^{2}}{\epsilon+\lambda_{1}^{*}
          + \cfrac{(v_{0}/2)^{2}\boldsymbol{k}^{2}}{\epsilon+\lambda_{2}^{*}
          + \cfrac{(v_{0}/2)^{2}\boldsymbol{k}^{2}}{\epsilon+\lambda_{3}^{*}+\ddots
           } }}
           },
\end{equation}
where the explicit dependence on the variable $\epsilon$ conveys that the Laplace transform \big[$f(\epsilon)=\int_{0}^{\infty}dt\, e^{-\epsilon t}f(t)$\big] has been 
carried out, and $\widetilde{p}_{0}^{(0)}(\boldsymbol{k})$ denotes the initial distribution $\widetilde{p}_{0}(\boldsymbol{k},t=0)$. The solution \eqref{p0} is {a 
generalization of the kind of solution obtained} in Ref. \cite{MehraeenPRE2008} for the {probability} distribution f{or} 
a {semiflexible polymer (modeled as an inextensible thread with a linear-elastic bending energy subjected to thermal fluctuations, known as }a wormlike chain{), 
that starts at the origin and ends at $\boldsymbol{x}$, independently of its orientation,} as has been also pointed out in Ref. \cite{DetcheverryEPJE2014} {in the context of 
random 
walks}.

In the present paper, the meaning of the solution \eqref{p0} can be elucidated after rewritten it as 
\begin{equation}
\label{p0-2}
 \widetilde{p}_{0}(\boldsymbol{k},\epsilon)=\frac{\widetilde{p}_{0}^{(0)}(\boldsymbol{k})}{\epsilon+(v_{0}/2)^{2}\, \boldsymbol{k}^{2}\,
\widetilde{\mathfrak{D}}(\boldsymbol{k},\epsilon)},
\end{equation}
or equivalently as
\begin{equation}
\label{p0-3}
 \epsilon\widetilde{p}_{0}(\boldsymbol{k},\epsilon)-\widetilde{p}_{0}^{(0)}(\boldsymbol{k})=-\left(\frac{v_{0}}{2}\right)^{2}\, \boldsymbol{k}^{2}\,
\widetilde{\mathfrak{D}}(\boldsymbol{k},\epsilon)\widetilde{p}_{0}(\boldsymbol{k},\epsilon),
\end{equation}
which can be recognized as the Fourier-Laplace transform of the spatially-non-local \emph{generalized diffusion equation},
\begin{equation}\label{Gral}
 \frac{\partial}{\partial t}p_{0}(\boldsymbol{x},t)=\left(\frac{v_{0}}{2}\right)^{2}\int d^{2}x^{\prime}\int_{0}^{t}ds\, 
\mathfrak{D}(\boldsymbol{x}-\boldsymbol{x}^{\prime},t-s){\nabla^{\prime}}^{2} p_{0}(\boldsymbol{x}^{\prime},s),
\end{equation}
introduced in Ref. \cite{KenkreSevilla2007} and {obtained} in the context of animal motion with internal states in {Ref.} \cite{giuggioli2009generalized}. 
{The integral over the 
spatial coordinates is computed over the whole two-dimensional plane.} The  \emph{connecting 
function} $\mathfrak{D}(\boldsymbol{x},t)$ is given explicitly in the Fourier-Laplace domain  by
\begin{equation}\label{Dke}
\widetilde{\mathfrak{D}}(\boldsymbol{k},\epsilon)=\cfrac{1}{\epsilon+\lambda_{1}
          + \cfrac{(v_{0}/2)^{2}\boldsymbol{k}^{2}}{\epsilon+\lambda_{2}
          + \cfrac{(v_{0}/2)^{2}\boldsymbol{k}^{2}}{\epsilon+\lambda_{3}+\ddots
           } } }+
	    \cfrac{1}{\epsilon+\lambda_{1}^{*}
          + \cfrac{(v_{0}/2)^{2}\boldsymbol{k}^{2}}{\epsilon+\lambda_{2}^{*}
          + \cfrac{(v_{0}/2)^{2}\boldsymbol{k}^{2}}{\epsilon+\lambda_{3}^{*}+\ddots
           } }}.
\end{equation}
\end{widetext}
We introduce the recursive relations
\begin{subequations}\label{RecursiveRels}
\begin{align}
  \Delta_{n}(\boldsymbol{k},\epsilon)&=\frac{1}{\epsilon+\lambda_{n+1}+(v_{0}/2)^{2}\boldsymbol{k}^{2}\Delta_{n+1}(\boldsymbol{k},\epsilon)},\\
  \overline{\Delta}_{n}(\boldsymbol{k},\epsilon)&=\frac{1}{\epsilon+\lambda_{-(n+1)}+(v_{0}/2)^{2}\boldsymbol{k}^{2}\overline{\Delta}_{n+1}(\boldsymbol{k},\epsilon)},
\end{align}
\end{subequations}
for $n\ge0$, to write Eq. \eqref{Dke} in a simplified form as
\begin{equation}\label{Dke-Recursive}
 \widetilde{\mathfrak{D}}(\boldsymbol{k},\epsilon)=\Delta_{0}(\boldsymbol{k},\epsilon)+\overline{\Delta}_{0}(\boldsymbol{k},\epsilon).
\end{equation}

In the asymptotic limit, i.e., in the long-time regime,  $\epsilon\rightarrow0$, and in the short-wave-vector limit, 
$k=\vert\boldsymbol{k}\vert\rightarrow0$, the 
connecting function 
is given by the zeroth order \emph{approximant}, $\widetilde{\mathfrak{D}}^{(0)}(\epsilon)$, obtained after evaluating 
$\widetilde{\mathfrak{D}}(\boldsymbol{k},\epsilon)$ at $\boldsymbol{k}=\boldsymbol{0}$, i.e.,
\begin{equation}\label{Approximant1-FL}
\widetilde{\mathfrak{D}}^{(0)}(\epsilon)\equiv\widetilde{\mathfrak{D}}(\boldsymbol{0},\epsilon)=\frac{1}{\epsilon+\lambda_{1}}+\frac{1}{
\epsilon+\lambda_{1}^{*}}.
\end{equation}
This implies a spatially-local connecting function, that exhibits oscillations of frequency $\Omega_{1}$, exponentially damped with relaxation time $\Gamma_{1}^ 
{-1}$, namely
\begin{equation}\label{Approximant1}
 \mathfrak{D}^{(0)}(\boldsymbol{x},t)=2\delta^{2}(\boldsymbol{x})e^{-\Gamma_{1}t}\cos\Omega_{1}t.
\end{equation}
With this approximation of the connecting function, we have that Eq. \eqref{Gral} can be rewritten in the form
\begin{equation}\label{TEChiral}
 \frac{\partial}{\partial t}p_{0}(\boldsymbol{x},t)=\frac{v_{0}^{2}}{2}\int_{0}^{t}ds\, e^{-\Gamma_{1}(t-s)}\cos\left[\Omega_{1}(t-s)\right]\, {\nabla}^{2} 
p_{0}(\boldsymbol{x},s),
\end{equation}
which corresponds to a generalization of the telegrapher equation in that it incorporates the effects of an effective torque that gives rise to circular motion of 
\emph{angular speed} $\Omega_{1}$. If $\Omega_{1}=0$, Eq. \eqref{TEChiral} reduces to the standard telegrapher's equation \cite{GoldsteinQJMAM1951}
\begin{equation}\label{TE}
 \frac{\partial^{2}}{\partial t^{2}}p_{0}(\boldsymbol{x},t)+\Gamma_{1}\frac{\partial}{\partial t}p_{0}(\boldsymbol{x},t)=\frac{v_{0}^{2}}{2}{\nabla}^{2} 
p_{0}(\boldsymbol{x},s),
\end{equation}
where the diffusion coefficient due to the persistence of the swimming direction, $D_{\text{pers}}=v_{0}^{2}/2\Gamma_{1}$, is apparent. {As before}, I 
identify 
$\Gamma_{1}^{-1}$ with the \emph{persistence time}. In the temporal asymptotic limit we have, from \eqref{p0}, that 
$\widetilde{p}_{0}(\boldsymbol{k},\epsilon)\sim\left[\epsilon+(v_{0}/2)^{2}\boldsymbol{k}^{2}\left(\lambda_{1}^{-1}+{\lambda_{1}^{*}}^{-1} \right)\right ] ^ { -1 }$, which 
can be inverted straightforwardly to the spatial and temporal variables to give the Gaussian $G_{D_{\text{eff}}}(\boldsymbol{x},t)$, i.e.,
\begin{equation}
 p_{0}(\boldsymbol{x},t)\sim \frac{1}{4\pi D_{\text{eff}}t}\exp\left\{-\frac{\boldsymbol{x}^{2}}{4D_{\text{eff}}t}\right\},
\end{equation}
where the effective diffusion coefficient, $D_{\text{eff}}$, due to active motion is defined by $D_{\text{eff}}=D_{\text{pers}}/(1+\Omega_{1}^{2}/\Gamma_{1}^{2})$, which 
reduces to $D_{\text{pers}}$ when $\Omega_{1}$ vanishes.

On the other hand, in the short time regime ($\vert\epsilon\vert\gg\vert\lambda_{n}\vert$ for all $n$), we have that 
$\widetilde{p}_{0}(\boldsymbol{k},\epsilon)$ can be approximated by 
$\widetilde{p}_{0}^{(0)}(\boldsymbol{k})\left[1/\epsilon-2(v_{0}/2)^{2}\boldsymbol{k}^{2}/\epsilon^{3}+\ldots\right]$. After taking the inverse Laplace transformation we 
obtain
\begin{equation}
 \widetilde{p}_{0}(\boldsymbol{k},t)\simeq\widetilde{p}_{0}^{(0)}(\boldsymbol{k})J_{0}(kv_{0}t),
\end{equation}
wich results from identifying the first two terms of the power series of the zeroth-order Bessel function of the first kind, $J_{0}(x)=1-(x/2)^{2}+\ldots$. For the initial 
distribution considered, we obtain the radial pulse:
\begin{equation}
 p_{0}(\boldsymbol{x},t)\simeq \frac{\delta(x-v_{0}t)}{2\pi x}
\end{equation}
wich propagates at speed $v_{0}$ free of the wakes exhibited by the solution of the approximated description given by the telegrapher's equation in 
the short time regime \cite{SevillaPRE2014}, where $x=\Vert\boldsymbol{x}\Vert$.

The next order approximant of $\widetilde{\mathfrak{D}}(\boldsymbol{k},\epsilon)$ is of particular interest since it leads to a connecting function coupled in the spatial 
and temporal variables. Some models of stochastic motion consider memory functions that couple space and time, as is the case for the family of stochastic motion known as 
\emph{L\'evy walks} --described within the formalism of \emph{continuous time random walks}-- where the transition probability density that connects two distinct points in 
space at different times is constrained by the condition that the walker moves at constant speed \cite{ZaburdaevRMP2015}. In our case the first order approximant, 
$\widetilde{\mathfrak{D}}^{(1)}(\boldsymbol{k},\epsilon)$, 
is obtained from $\widetilde{\mathfrak{D}}(\boldsymbol{k},\epsilon)$ after evaluating $\Delta_{1}(\boldsymbol{k},t)$ and 
$\overline{\Delta}_{1}(\boldsymbol{k},\epsilon)$ at $\boldsymbol{k}=\boldsymbol{0}$, which leads to 
\begin{multline}
 \widetilde{\mathfrak{D}}^{(1)}(\boldsymbol{k},\epsilon)=\cfrac{1}{\epsilon+\lambda_{1}
          + \cfrac{(v_{0}/2)^{2}\boldsymbol{k}^{2}}{\epsilon+\lambda_{2}
          } }+\cfrac{1}{\epsilon+\lambda_{1}^{*}
          + \cfrac{(v_{0}/2)^{2}\boldsymbol{k}^{2}}{\epsilon+\lambda_{2}^{*}
           }}.
\end{multline}

In the time regime for which $\vert\epsilon\vert\ll\vert\lambda_{2}\vert$, an explicit expression for $\mathfrak{D}^{(1)}(\boldsymbol{x},t)$ in spatial and 
temporal 
coordinates is obtained, namely
\begin{multline}\label{Approximant2}
\mathfrak{D}^{(1)}(\boldsymbol{x},t)=2e^{-\Gamma_{1}t}G_{v_{0}^{2}/4\Gamma_{2}}(\boldsymbol{x},t)\times\\
\left\{\cos\left[\Omega_{1}t\left(1+\frac{
\Omega_ { 2 } } { \Omega_ {1}} \frac{\boldsymbol{x}^{2}}{v_{0}^{2}t^{2}}\right)\right]\right.+\\
\left.\frac{\Omega_{2}}{\Gamma_{2}}\sin\left[\Omega_{1}t\left(1+\frac{\Omega_{2}}{
\Omega_ {1}} \frac{\boldsymbol{x}^{2}}{v_{0}^{2}t^{2}}\right)\right]\right\}.
\end{multline}
Due to the explicit appearance of the Gaussian $G_{v_{0}^{2}/4\Gamma_{2}}(\boldsymbol{x},t)$, the connecting function \eqref{Approximant2} gives a major contribution to those 
spatial positions $\boldsymbol{x}$, $\boldsymbol{x}^{\prime}$, whose separation is less or of the order of  the distance $\sqrt{v_{0}^{2}t/\Gamma_{2}}$, and decays quickly to 
zero for pairs of points whose distance is larger than this.  It is expected that the Gaussian nonlocality of \eqref{Approximant2}, is a consequence of the 
approximation made, and that a connecting function that vanishes for pair of points  whose distance is larger than $v_{0}t$ is more appropriate. Note that \eqref{Approximant2} 
reduces to the long-time approximation given by \eqref{Approximant1}, by taking the limit $\Omega_{2},\Gamma_{2}\rightarrow0$. 

\subsection{The connecting function $\widetilde{\mathfrak{D}}(\boldsymbol{k},\epsilon)$ and the moments of $p_{0}(\boldsymbol{x},t)$ }

For the initial condition considered, we have that the solution given in Eq. \eqref{p0-2} is a rotationally symmetric function 
that depends solely on $\boldsymbol{k}^{2}$, and we simply write $\widetilde{p}_{0}(k,\epsilon)$. Likewise, we can write 
$p_{0}(\boldsymbol{x},\epsilon)=(2\pi)^{-1}p_{0}(x,\epsilon)$, where $x$ denotes the magnitude of $\boldsymbol{x}$, and the explicit appearance of the Laplace variable 
$\epsilon$ indicates that the Laplace transform is considered. The mentioned rotational symmetry allows to write {\eqref{p0FourierLaplace} as
\begin{equation}
  \widetilde{p}_{0}(k,\epsilon)=\frac{1}{2\pi}\int_{0}^{\infty} dx\, x\,p_{0}(x,\epsilon)J_{0}(kx),
\end{equation}
i.e., $\widetilde{p}_{0}(k,\epsilon)=\langle J_{0}(kx)\rangle_{\text{rad}}$,
}
where $\langle z[x(\epsilon)]\rangle_{\text{rad}}$ denotes the average of $z(x)$ over the radial distribution $x\,p_{0}(x,\epsilon)$, {thus, after use of the power 
series representation of the Bessel function $J_{0}(x)=\sum_{l=0}^{\infty}\bigl[(-1)^{l}/(l!)^{2}\bigr](x/2)^{2l}$ we have that}
\begin{equation}
\widetilde{p}_{0}(k,\epsilon)=\frac{1}{2\pi}\sum_{n=0}^{\infty}\frac{(-1)^{n}}{(n!)^{2}}\frac{k^{2n}}{2^{2n}}\langle x^{2n}(\epsilon)\rangle_{\text{rad}},
\end{equation}
{where $\langle x^{2n}(\epsilon)\rangle_{\text{rad}}$ are} the rotationally symmetric moments given by
\begin{equation}
 \langle x^{2n}(\epsilon)\rangle_{\text{rad}}=\int_{0}^{\infty}dx\, x^{2n}\, x\, p_{0}(x,\epsilon).
 \end{equation}
{These can also be obtained directly from $\widetilde{p}_{0}(k,\epsilon)$, if this is known,} from the formula
\begin{equation}\label{RadialMoments}
 \langle x^{2n}(\epsilon)\rangle_{\text{rad}}=2\pi\frac{(-1)^{n}n!\,2^{n}}{(2n-1)!!}\left.\frac{d^{2n}}{dk^{2n}}\widetilde{p}_{0}(k,\epsilon)\right\vert_{k=0}.
\end{equation}

\subsubsection{The mean-square displacement}
The mean-square displacement is defined by $\langle\boldsymbol{x}^{2}(t)\rangle$, which coincides with $\langle x^{2}(t)\rangle_{\text{rad}}$. It follows 
straightforwardly from \eqref{RadialMoments} {with $n=1$}, that the Laplace transform of the mean-square displacement is given by
{\begin{equation}\label{MSDFormula}
 \langle\boldsymbol{x}^{2}(\epsilon)\rangle=-4\pi\left.\frac{d^{2}}{dk^{2}}\widetilde{p}_{0}(k,\epsilon)\right\vert_{k=0}.
\end{equation}
By substitution of expression \eqref{p0-2} for {the probability density independent of the direction of motion,} $\widetilde{p}_{0}(k,\epsilon)$, in the last equation, 
with the initial condition $\widetilde{p}_{0}^{(0)}(k)=1/2\pi$, we have 
that (see the Appendix \ref{appendix:MSD})}
\begin{equation}\label{MSDLaplace}
\langle\boldsymbol{x}^{2}(\epsilon)\rangle=\left.\frac{v_{0}^{2}}{\epsilon^{2}}\widetilde{\mathfrak{D}}(\boldsymbol{k},\epsilon)\right\vert_{k=0}=\frac{v_{0}^{2}}{
\epsilon^{2}}\widetilde{\mathfrak{D}}^{(0)}(\epsilon),
\end{equation}
where
$\left.\widetilde{\mathfrak{D}}(\boldsymbol{k},\epsilon)\right\vert_{k=0}$ corresponds to the zeroth-order approximant
$\widetilde{\mathfrak{D}}^{(0)}(\epsilon)$ of  $\widetilde{\mathfrak{D}}(\boldsymbol{k},\epsilon)$, 
given in \eqref{Approximant1-FL}. After inverting the Laplace transform, the exact time dependence of the mean-square displacement is given by
\begin{multline}\label{MSD}
 \langle\boldsymbol{x}^{2}(t)\rangle=4\frac{D_{\text{eff}}}{\Gamma_{1}}\, 
\left[\Gamma_{1}t-\frac{1-\frac{\Omega_{1}^{2}}{\Gamma_{1}^{2}}}{1+\frac{\Omega_{1}^{2}}{\Gamma_{1}^{2}}}\left(1-e^{-\Gamma_{1}t}\cos\Omega_{1}t\right)\right.\\
\left.-\frac{2\frac{\Omega_{1}}{\Gamma_{1}}}{1+\frac{\Omega_{1}^{2}}{\Gamma_{1}^{2}}}e^{-\Gamma_{1}t}\sin\Omega_{1}t\right],
\end{multline}
which reduces to the well-known expression
\begin{equation}
 \langle\boldsymbol{x}^{2}(t)\rangle=\frac{2v_{0}^{2}}{\Gamma_{1}^{2}}\left[\Gamma_{1}t-\left(1-e^{-\Gamma_{1}t}\right)\right],
\end{equation}
when $\Omega_{1}$ vanishes.
\begin{figure}[t]
\includegraphics[width=\columnwidth]{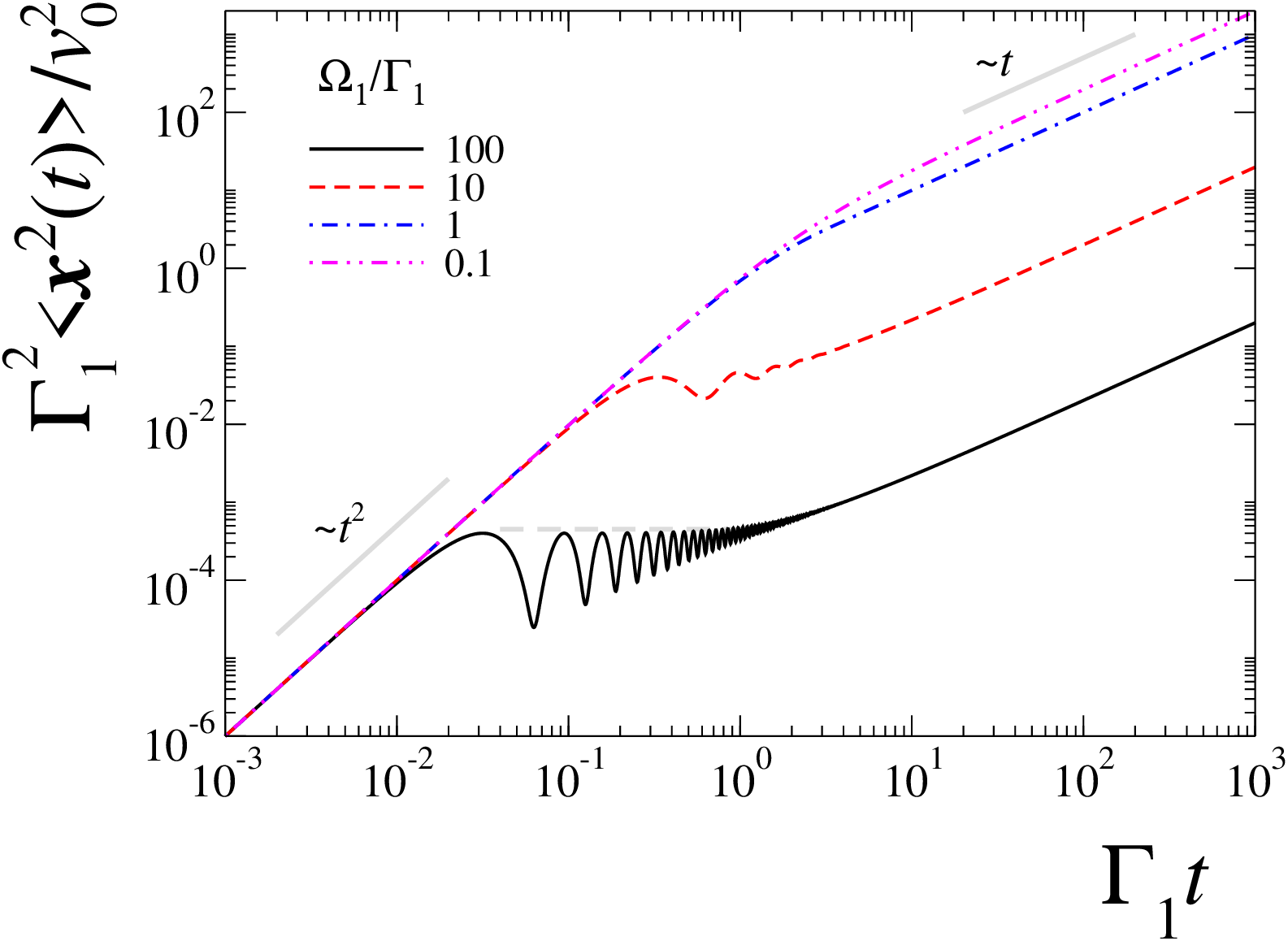}
\caption{(Color online) Dimensionless mean-square displacement $\Gamma_{1}^{2}\langle\boldsymbol{x}^{2}(t)\rangle/v_{0}^{2}$ as function of the dimensionless time 
$\Gamma_{1}t$ for different values of the ratio $\Omega_{1}/\Gamma_{1}$, namely, 0.1, 1, 10, 100.} 
\label{FIG_MSD}
\end{figure}

{It is noticed,} from expression \eqref{MSD}, {that} the regime for which the particle motion is dominantly ballistic, $\langle\boldsymbol{x}^{2}(t)\rangle\rightarrow 
v_{0}^{2}t^{2}$, is obtained in the short-time regime, $\Gamma_{1}t\ll1$, for arbitrary $\Omega_{1}$ (see Fig. \ref{FIG_MSD}). In contrast, in the long-time 
regime, $\Gamma_{1} t\gg1$, we get 
the standard linear dependence in time of the mean-square displacement
\begin{equation}\label{MSDLongTime}
 \langle\boldsymbol{x}^{2}(t)\rangle\sim 4D_{\text{eff}}\, t,
\end{equation}
with $D_{\text{eff}}$,  given as before, as $D_{\text{pers}}/(1+\Omega_{1}^{2}/\Gamma_{1}^{2})$. As is well-known 
\cite{LarraldePRE1997, WeberPRE2011}, the effective diffusion coefficient reaches its maximum value 
\begin{equation}\label{DeffMax}
D_{\text{eff}}^{*}=v_{0}^{2}/4\Omega_{1} 
\end{equation}
at the ratio $\Omega_{1}/\Gamma_{1}=1$. It can be clearly noticed from Eq. \eqref{MSD}, that
the time dependence of the mean-square displacement depends only on the ratio $\Omega_{1}^{2}/\Gamma_{1}^{2}$, whose explicit value depends on the particular transition 
probability 
density $\widetilde{Q}(\varphi)$. Thus, the crossover from the ballistic regime to the normal diffusion one is sensitive to the particular details of the pattern of active 
motion, entailed in $\Omega_{1}/\Gamma_{1}$ through $\widetilde{Q}(\varphi)$ as is shown in Fig. \ref{FIG_MSD}. For large values of the ratio $\Omega_{1}/\Gamma_{1}$, the 
particle get self-trapped in the intermediate-time regime due to the circular motion induced by the particular choice of $\widetilde{Q}(\varphi)$, and revealed by the 
corresponding oscillations of the mean-square displacement (see the solid-black line in Fig. \ref{FIG_MSD} for $\Omega_{1}/\Gamma_{1}=100$).

\subsubsection{\label{subsect:kurtosis}The kurtosis}
The non-Gaussian feature of the probability density $p_{0}(\boldsymbol{x},t)$ can be characterized by its kurtosis $\kappa$, which as a matter of convenience, the 
definition given by Mardia \cite{Mardia74p115} is used, namely
\begin{equation}
\kappa(t) =\left\langle \left[\bigl(\boldsymbol{x}(t)-\langle \boldsymbol{x}(t)\rangle\bigr)\Sigma^{-1}\bigl(\boldsymbol{x}(t)-\langle 
\boldsymbol{x}(t)\rangle\bigr)^{\text{T}}\right]
^{2}\right\rangle ,  \label{kur}
\end{equation}%
where $\boldsymbol{x}^{\text{T}}$ denotes the transpose of the vector $\boldsymbol{x}$ and $\Sigma $ is the $2\times2$ matrix defined by the average of the dyadic product 
$\bigl(\boldsymbol{x}(t)-\langle \boldsymbol{x}(t)\rangle\bigr)^{\text{T}}\cdot \bigl(\boldsymbol{x}(t) -\langle \boldsymbol{x}(t)\rangle\bigr).$ For the circularly 
symmetric case, the one considered in 
this paper, Eq. (\ref{kur}) reduces to
\begin{equation}\label{ksim}
\kappa(t) =4\frac{\langle x^{4}(t)\rangle _{\text{rad}}}{\langle x^{2}(t)\rangle _{\text{rad}}^{2}}. 
\end{equation}
From Eq. \eqref{RadialMoments} we have that the Laplace transform of the time dependence of the fourth-moment is given by
\begin{equation}\label{FourthMomentLaplace}
 \langle x^{4}(\epsilon)\rangle _{\text{rad}}=\frac{4v_{0}^{4}}{\epsilon^{3}}\left[\widetilde{\mathfrak{D}}(\boldsymbol{k},\epsilon)\right]^{2}_{k=0}
 -\frac{8v_{0}^{2}}{\epsilon^{2}}\left[\frac{\partial^{2}}{\partial k^{2}}\widetilde{\mathfrak{D}}(\boldsymbol{k},\epsilon)\right]_{k=0}.
\end{equation}
Notice that the first term in the last equation depends solely on $\lambda_{1}$, $\lambda_{1}^{*}$ 
{since $\left[\widetilde{\mathfrak{D}}(\boldsymbol{k},\epsilon)\right]^{2}_{k=0}=\left[\widetilde{\mathfrak{D}}^{(0)}(\epsilon)\right]^{2}=\left[(\epsilon+\lambda_{1})^{-1}
+(\epsilon+\lambda_{1}^{*})^{-1}\right]^{2}$} , while the second term carries 
information about $\lambda_{2}$, $\lambda_{2}^{*}$ {(and therefore about $\Gamma_{2}$ and $\Omega_{2}$)} since, as can be shown straightforwardly from \eqref{Dke} and 
\eqref{RecursiveRels} (see the Appendix \ref{appendix:MSD}), 
\begin{multline}\label{2derivativeDke}
 \left.\frac{\partial^{2}}{\partial 
k^{2}}\widetilde{\mathfrak{D}}(\boldsymbol{k},t)\right\vert_{k=0}=-\frac{v_{0}^{2}}{2}\left[\frac{1}{(\epsilon+\lambda_{1})^{2}(\epsilon+\lambda_{2})}+\right.\\
\left.\frac{1}{(\epsilon+\lambda_{1}^{*})^{2}(\epsilon+\lambda_{2}^{*})}\right].
\end{multline}
Thus the fourth moment in Laplace domain is explicitly given by
\begin{multline}\label{4thMoment-L}
 \langle 
x^{4}(\epsilon)\rangle=\frac{4v_{0}^{4}}{\epsilon^{2}}\left[\frac{1}{\epsilon}\left(\frac{1}{\epsilon+\lambda_{1}}+\frac{1}{\epsilon+\lambda_{1}^{*}}\right)^{2}+\right.\\
\left.\frac{1}{(\epsilon+\lambda_{1})^{2}(\epsilon+\lambda_{2})}+\frac{1}{(\epsilon+\lambda_{1}^{*})^{2}(\epsilon+\lambda_{2}^{*})} \right].
\end{multline}

The general explicit time dependence of the fourth moment is too involved to be discussed at this point. Besides, the values of $\Gamma_{1}$, 
$\Omega_{1}$, $\Gamma_{2}$ and $\Omega_{2}$ are not independent among them, but they are related through the transition probability density $\widetilde{Q}(\varphi)$. Thus, an 
analysis of the time dependence of the kurtosis is presented in the next section for particular cases of $\widetilde{Q}(\varphi)$. Notwithstanding this, the short- and 
long-time regimes can be discussed straightforwardly.

In the long-time regime ($\vert\epsilon\vert\ll\vert\lambda_{1}\vert,\,\vert\lambda_{2}\vert$), the second and third terms in the squared brackets of Eq. 
\eqref{4thMoment-L} can be neglected with respect to the first one, and thus, it is the first term that mainly contributes in the long-time regime. In such 
regime the fourth moment is independent 
of $\lambda_{2}$ and $\lambda_{2}^{*}$, and the 
inversion of the Laplace transform can be done straightforwardly, which gives
\begin{equation}
 \langle x^{4}(t)\rangle\sim8\frac{v_{0}^{4}\Gamma_{1}^{2}}{\left(\Gamma_{1}^{2}+\Omega_{1}^{2}\right)^{2}}t^{2},
\end{equation}
and from this, we deduce that $\kappa\sim8$, which uniquely characterizes the two-dimensional Gaussian distribution. Unlike this case, in the 
short-time 
regime ($\vert\epsilon\vert\gg\vert\lambda_{1}\vert,\,\vert\lambda_{2}\vert$), we have that all the terms in Eq. \eqref{4thMoment-L} contribute, and such an expression 
reduces to $4!v_{0}^{4}/\epsilon^{5}$, which can be 
inverted to give $v_{0}^{4}t^{4}$ (independent of $\lambda_{1}$, $\lambda_{2}$ and their complex conjugates), and thus $\kappa\simeq 4$ which characterizes the 
distortionless propagation of the sharp pulse $\delta(x-v_{0}t)/(2\pi x)$ \cite{SevillaPRE2014,SevillaPRE2015}.

The effects of $\lambda_{2}$, $\lambda_{2}^{*}$ can be observed only in the intermediate-time regime, where the particle positions distribution suffers of the important 
effects of persistence, as has been anticipated in Sect. \ref{SubSectIIIA} as is discussed in the following sections.

\section{\label{SectParticularCases}Persistence time, natural period of rotation and other time-scales}

As has been already introduced in Sec. \ref{SectGralSolution}, the \emph{persistence 
time} of the swimming direction, $\Gamma_{1}^{-1}$, and the \emph{natural period} of the circular motion, $\Omega_{1}^{-1}$, correspond to  the relevant time-scales  
that define the diffusive regime of the active motion [see Eq. 
\eqref{MSDLongTime}]. $\Gamma_{1}^{-1}$ is closely related to the persistence time introduced by Wu \emph{et al.} in Ref. \cite{WuEcoMod2000}, and by Bartumeus \emph{et al.} 
in Ref. \cite{BartumeusJTB2008} in the modeling and analysis of animal motion in two dimensions as correlated random walks. All the other time-scales that appear in the 
present analysis [see, for instance, the expansion Eq. \eqref{Pexpansion2}], namely, $\Gamma_{n}^{-1}$, 
$\Omega_{n}^{-1}$, with $n>1$, determine precisely the statistical properties of active motion at all time regimes. These depend on the particular 
choice of the scattering-angle distribution $\widetilde{Q}(\varphi)$. 

The simplest scattering-angle distribution may correspond to the case when $\widetilde{Q}(\varphi)$ is uniform in $[-\pi,\pi]$, i.e., $\widetilde{Q}(\varphi)=(2\pi)^{-1}$. 
This has been used to model the paradigmatic two-dimensional pattern of active motion called \emph{run-and-tumble} 
\cite{SchnitzerPRE1993,PorraPRE1997,Solon2015EPJST2015,HancockPRE2015}, for 
which $\Gamma_{n}=\Lambda$ for all $n$, i.e., $\Lambda{^{-1}}$ is the unique time-scale that defines the dynamics of the swimming direction, meaning that all Fourier 
modes in the series \eqref{Pexpansion} decay at the same pace $\Lambda$.

Moreover, many scattering-angle distributions can be built on by \emph{wrapping} out a standard single-variate distribution, $\rho(\eta)$, with 
support on the interval $(-\infty,\infty)$, to the unitary circle, namely 
\begin{equation}\label{Qwrapped}
\widetilde{Q}_{\text{wr}}(\varphi)=\int_{-\infty}^{\infty}d\eta\, \rho(\eta)\sum_{m=-\infty}^{\infty}\delta(\eta-\varphi+2\pi m).
\end{equation}
One important set of scattering-angle distributions obtained in this manner, is got from the well-known L\'evy $\alpha$-stable distributions with index $0<\alpha\le2$, 
$\rho_{\alpha;\sigma,\phi,\beta}(\eta)$, whose characteristic function is given by
\begin{equation}\label{LevyDistributions}
 \hat{\rho}_{\alpha;\sigma,\phi,\beta}(\kappa)=\exp\left\{i\kappa\phi-\vert\sigma\kappa\vert^{\alpha}\bigl(1-i\beta\text{sign}(\kappa)\bigr)\Phi\right\},
\end{equation}
being  $\sigma>0$ the width, $\phi$ the mode, and $\beta$ the skewness, 
$\Phi$ equals $\tan(\pi\alpha/2)$ if $\alpha\neq1$ and $-2\ln\vert \kappa\vert/\pi$ if $\alpha=1$. The cases $\alpha=2$, $\alpha=1$, with $\beta=0$; and $\alpha=1/2$, 
with $\beta=1$, 
are of interest, since these cases correspond to the wrapped Gaussian and the wrapped Lorentz (Cauchy) distribution{s in the first cases} and 
to the wrapped L\'evy distribution in the second one. For {the L\'evy $\alpha$-stable distributions 
\eqref{LevyDistributions},} it is possible to obtain explicit expressions for $\Gamma_{n}$ and $\Omega_{n}$, we have for $n>1$ that 
\begin{subequations}\label{ScalesStableDist}
 \begin{align}
    \Gamma_{n}&=\Lambda\left[1-e^{-(\sigma n)^{\alpha}}\cos\Bigl(n\phi+(\sigma n)^{\alpha}\beta\Phi\Bigr)\right],\\
    \Omega_{n}&=\Lambda e^{-(\sigma n)^{\alpha}}\sin\Bigl(n\phi+(\sigma n)^{\alpha}\beta\Phi\Bigr).
 \end{align}
\end{subequations}

Another important family of scattering-angle distributions is the one given by the angle distribution of  Jones and Pewsey, 
$\widetilde{Q}_{\text{JP},\sigma,\phi,\psi}(\varphi)$ \cite{JonesCircleDistributions2005}, with parameters: $\sigma>0$, $\phi$, and $\psi\in(-\infty,\infty)$, which correspond 
respectively to the distribution width, the location of the unique mode,  and the shape parameter. It has the explicit representation
\begin{equation}\label{JonesPewseyDistriution}
 \widetilde{Q}_{\text{JP},\sigma,\phi,\psi}(\varphi)=\frac{\bigl[\cosh(\sigma\psi)+\sinh(\sigma\psi)\cos(\varphi-\phi)\bigr]^{1/\psi}}{2\pi 
P_{1/\psi}\left[\cosh(\sigma\psi)\right]},
\end{equation}
where $P_{\gamma}(z)$ is the associated Legendre function of the first kind of degree $\gamma$. The distribution \eqref{JonesPewseyDistriution}
contains as particular cases \cite{JonesCircleDistributions2005}: the angle distribution of von Misses ($\psi=0$)
\begin{equation}
 \widetilde{Q}_{\text{vM}}(\varphi)=\frac{e^{\kappa\cos\varphi}}{2\pi I_{0}(\kappa)},
\end{equation}
the cardioid distribution ($\psi=-1$)
\begin{equation}
 \widetilde{Q}_{\text{CD}}(\varphi)=\frac{1}{2\pi}\bigl(1+\tanh(\kappa)\cos\varphi\bigr),
\end{equation}
and the wrapped Cauchy distribution ($\psi=1$)
\begin{equation}
 \widetilde{Q}_{\text{C}}(\varphi)=\frac{1}{2\pi}\frac{1-\tanh^{2}(\frac{\kappa}{2})}{1+\tanh^{2}(\frac{\kappa}{2})-2\tanh(\frac{\kappa}{2})\cos\varphi}.
\end{equation}
The distribution \eqref{JonesPewseyDistriution} has also been used in the analysis of two-dimensional correlated random walks \cite{BartumeusJTB2008}.

Although the number of possibilities to make a choice of the turning-angle distribution is vast, we focus our analysis on two wide-enough classes of the 
scattering functions: a class of unimodal distributions and one of bimodal ones. Subclasses will be defined by features such as the symmetry with 
respect the turning angle zero, and will endow with specific properties, to the quantities $\Gamma_{n}$ and $\Omega_{n}$.

\subsection{Unimodal angular distributions}
Lets first consider the case of unimodal distributions, which splits into two wide categories: the symmetric scattering-angle distributions around the 
instantaneous swimming direction, i.e., the distributions $\widetilde{Q}(\varphi)$  for whose single one mode is centered about $\varphi=0,$ or $\pm \pi$; and the asymmetric 
ones, whose mode is located at some value on the interval $[-\pi,\pi]$, except $0$ or $\pi$. 

\subsubsection{Symmetric scattering-angle distributions}

Smooth-enough unimodal distributions, $\widetilde{Q}_{\text{S}}(\varphi)$, that are symmetrically distributed around the mode  $\phi=0$ or around the mode $\phi=\pm\pi$, are 
of great interest since there is a variety of biological organisms, and artificially designed particles too, that follow this pattern (strategy) of motion 
(see for instance Ref. \cite{BechingerRMP2016} for a variety of microswimmers---like Janus particles, \emph{E. coli}, etc---that can be described by a dynamics 
of the particle reorientation along the forward direction of motion, and Refs. \cite{ZhangPNAS2010,LiuPRL2019} for organisms that exhibits dynamics of the particle 
reorientation along the backward direction of motion). When the 
scattered angle is distributed around {the forward direction of motion (}$\phi=0${)}, the motion is highly persistent, and it is perhaps the most ubiquitous pattern 
of active motion observed. This type of dynamics is widely known as rotational diffusion dynamics \cite{BechingerRMP2016}. On the contrary, 
motion becomes highly antipersistent if the distribution of scattered angles is centered  around $\pm\pi$, a pattern of motion known as \emph{run-and-reverse}, exhibited by 
the bacteria \emph{Myxococcus xanthus} \cite{LiuPRL2019} and a variety of other microorganisms \cite{[{}][{and references therein}]{GrossmannNJP2016}}. In both cases we 
have that $\Omega_{n}=0$ for all $n$, since $\widetilde{Q}(\varphi)$, being an even function of $\varphi$ or $\varphi\pm\pi$, makes 
$\langle\sin\varphi\rangle_{\widetilde{Q}}$ to vanish in this case. 

\begin{figure}[h]
 \includegraphics[width=\columnwidth, clip]{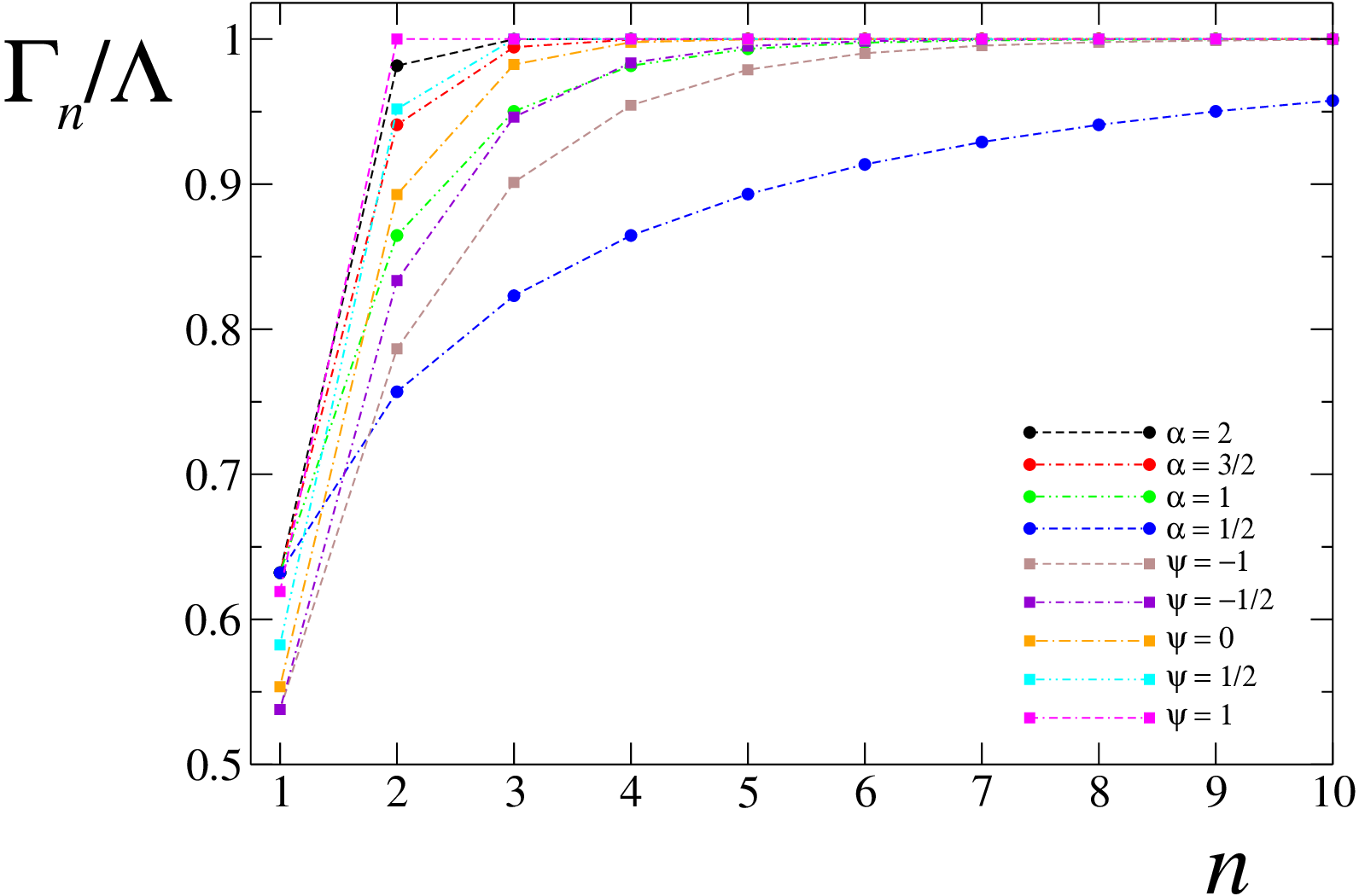}
 \caption{(Color online) The first 10 values of $\Gamma_{n}/\Lambda$ are shown for different unimodal distributions of scattered angles centered at the forward direction of 
motion. For the L\'evy alpha-stable distributions wrapped to the circle, $\widetilde{Q}_{\text{wr}}(\varphi)$, $\alpha=2$ (wrapped Gaussian), 3/2, 1 (wrapped Lorentz), and 
1/2 were chosen, all with parameters $\sigma=1$, $\beta=0$. For the Jones and Pewsey distributions of scattered angles, $\widetilde{Q}_{\text{JP},\sigma,\phi,\psi}$ the 
values $\sigma=1$, $\phi=0$, and $\psi=-1$ (cardioid), $-1/2$, 0 (von Misses), 1/2, and 1 (wrapped Cauchy) were chosen. Notice the saturation value 
$\Gamma_{n}/\Lambda=1$, which is half the maximum value allowed (see text in Sec. \ref{SectGralSolution}). }
 \label{Gamma_n_UnimodalSymmetricDist}
\end{figure}

\begin{figure}[h]
 \includegraphics[width=0.982\columnwidth]{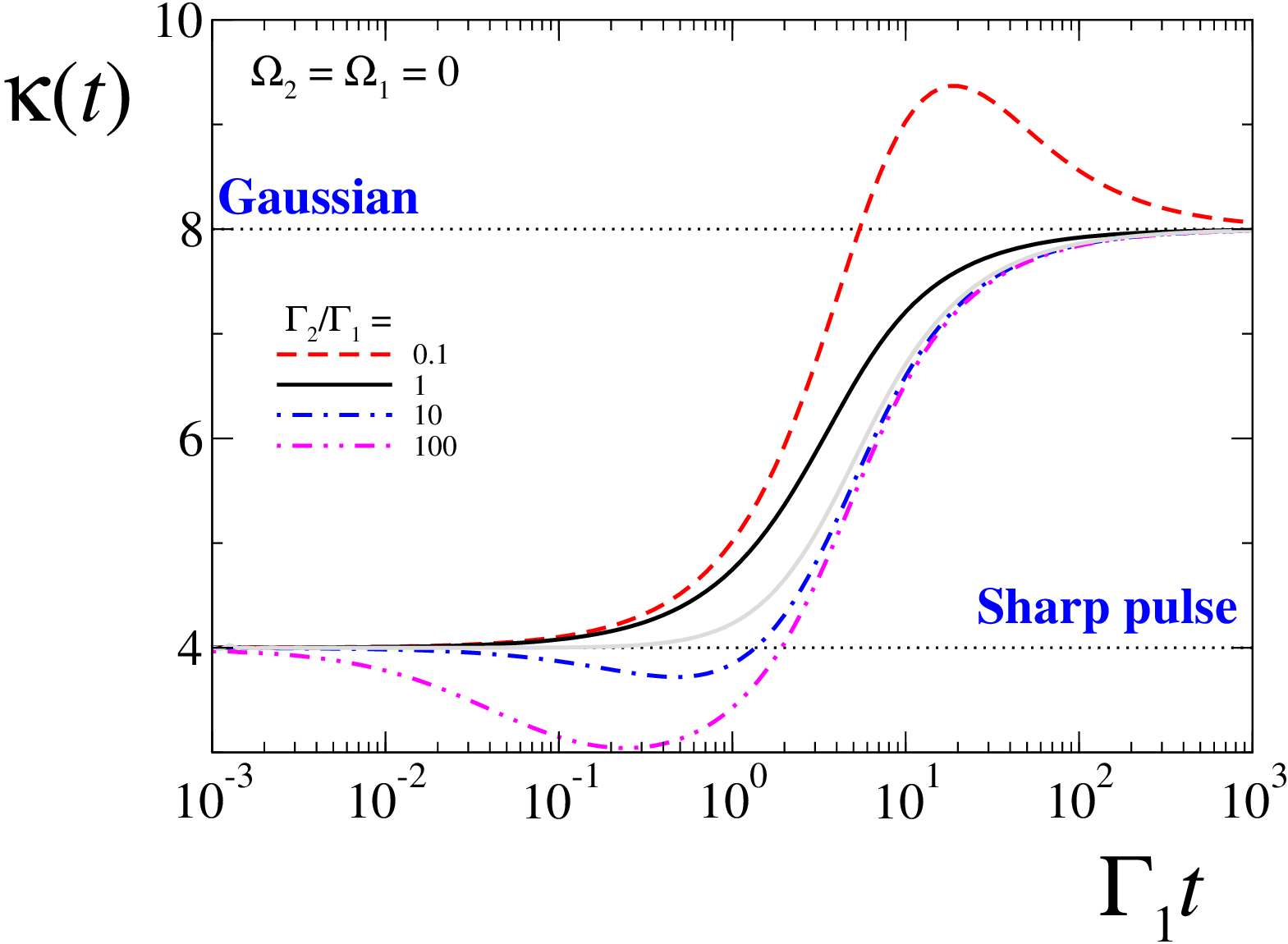}
 \caption{(Color online) Kurtosis as function of the dimensionless time $\Gamma_{1}t$ for symmetric scattering-angle distribution $\widetilde{Q}_{\text{S}}(\varphi)$. The 
values of the ratio $\Gamma_{2}/\Gamma_{1}=0.1$ (dashed-red line), 1 (solid-black line given by Eq. \eqref{KurtosisRTP}), 10 (dashed-dotted-blue line), and 
100 (dashed-double-dotted-magenta line) have been considered. Horizontal thin-dotted lines mark the values $\kappa=8$ and $4$ that correspond to the cases for 
which the probability density $p_{0}(\boldsymbol{x},t)$ is Gaussian in the long-time regime ($\kappa=8$), and a sharp pulse that propagates with speed $v_{0}$ ($\kappa=4$), 
respectively. The solid-golden line corresponds to the time dependence of the kurtosis for active Brownian motion with rotational diffusion constant equal to $\Gamma_{1}$.}
 \label{KurtosisQSym}
\end{figure}

When the direction of motion is frequently scattered forwardly, i.e., around the instantaneous direction of motion, the mode of $\widetilde{Q}_{\text{S}}(\varphi)$ 
is located at $\phi=0$ and it can be shown that $0<\Gamma_{n}\le\Gamma_{m}$ whenever $n<m$ (see Fig. \ref{Gamma_n_UnimodalSymmetricDist} for some specific symmetric 
unimodal distributions).  Particularly, we have that $\Gamma_{2}/\Gamma_{1}\ge1$, {or $\Gamma_{2}^{-1}\le\Gamma_{1}^{-1}$,} and  the effects of 
this are revealed by the kurtosis during times before the persistence time, $\Gamma_{1}t\lesssim1$, (see dashed-dotted-blue and dashed-double-dotted-magenta lines in 
Fig. \ref{KurtosisQSym} for $\Gamma_{2}/\Gamma_{1}=10$ and 100, respectively). 
In such a period of time, the initial sharp pulse diminishes from its characteristic value $\kappa=4$ giving rise to wakes, which are 
characteristic of wavelike propagation in two dimensions (see Ref. \cite{SevillaPRE2014}). This can be appreciated in Fig. \ref{KurtosisQSym}, where the transit 
from the initial sharp pulse ($\kappa=4$) to the Gaussian distribution ($\kappa=8$) is not monotonic.

On the contrary, the inequality $\Gamma_{2}/\Gamma_{1}\le1$ is satisfied for scattered angles that frequently occur around the contrary direction to the instantaneous 
direction of motion, i.e., when the mode $\phi$ is located at $\pi$. Thus, these effects are revealed in the kurtosis at times $t$ for which $\Gamma_{1}t\gtrsim1$ (see 
dashed-red line in Fig. \ref{KurtosisQSym}). There are no wakes in the propagation pulse in short-time regime, but now the distribution becomes conspicuously leptokurtic 
(more acute than Gaussian for which $\kappa>8$) asymptotically tending to the Gaussian.

These inequalities give a clear insight of the role of the properties of the scattering-angles distribution{, $\widetilde{Q}(\varphi)$,} on the  time evolution of the 
\textquotedblleft shape\textquotedblright of $p_{0}(\boldsymbol{x},t)$, characterized by its kurtosis $\kappa(t)$. 

The case $\Gamma_{2}/\Gamma_{1}=1$ is of some interest and leads to a simple expression for the kurtosis, namely,
\begin{equation}\label{KurtosisRTP}
 \kappa(t)=24\frac{1-\Gamma t+\Gamma^{2}t^{2}/3-e^{-\Gamma t}+\Gamma^{2}t^{2}e^{-\Gamma t}/6}{[\Gamma t-(1-e^{-\Gamma t})]^{2}},
\end{equation}
where we have written $\Gamma_{1}=\Gamma_{2}=\Gamma$. This particular case has as an instance, the well-known pattern of active motion called \emph{run-and-tumble}, for 
which 
$\widetilde{Q}_{\text{S}}(\varphi)=(2\pi)^{-1}$ and therefore $\Gamma_{n}=\Lambda$ for all $n$. Notice that the \textquotedblleft shape\textquotedblright of 
$p_{0}(\boldsymbol{x},t)$ changes from the initial sharp pulse to the Gaussian distribution in a monotonic way, as can be deduced from the monotonic-nondecreasing time 
dependence of the kurtosis ($4\le\kappa(t)\le8$ at all instants). This monotonic growth is representative of many patterns of active motion for which the direction of motion 
is slightly scattered from  the instantaneous one. For comparison purposes, I have included in Fig. \ref{KurtosisQSym} the time dependence of the 
kurtosis for active Brownian motion \cite{SevillaPRE2014} (solid-golden line), for which the persistence of the direction of motion is lost by rotational 
diffusion.
 
\subsubsection{Asymmetric scattering-angle distributions}
Unimodal distributions that consider a frequent scattering of the swimming direction to{wards} directions of motion different from the forward one, or the reverse one, 
i.e., those that have a mode at angles $\phi\neq0$, $\pi$, lead naturally to \emph{circular motion}. This is expected even in the case of forward or 
reverse scattering, however circular motion emerges as consequence of the skewness of the distribution $\widetilde{Q}(\varphi)$, i.e., when angles are more frequently 
scattered clockwise or anticlockwise. These statistical considerations allow to describe the motion of \emph{circular swimmers}, which are ubiquitous in nature and 
have been observed in a variety of biological organisms and of artificially designed swimmers 
\cite{Crenshaw1BullMathBio1993,LaugaBioPhysJ2006,ShenoyPNAS2007,SchmidtEBioPhysJ2008,FriedrichNJP2008,MarinePRE2013,KummelPRL2013,TakagiPRL2013,
Gomez-SolanoPRL2016,NarinderPRL2018}, these swimmers have been of theoretical interest leading to diverse models that describe their motion 
\cite{VanTeeffelenPRE2008,FriedrichPRL2009,OhtaPRL2009,WeberPRE2011,WittkowskiPRE2012,Ledesma-AguilarEPJE2012,LowenEPJST2016,KurzthalerSoftMatter2017}.

\begin{figure}
 \includegraphics[width=\columnwidth]{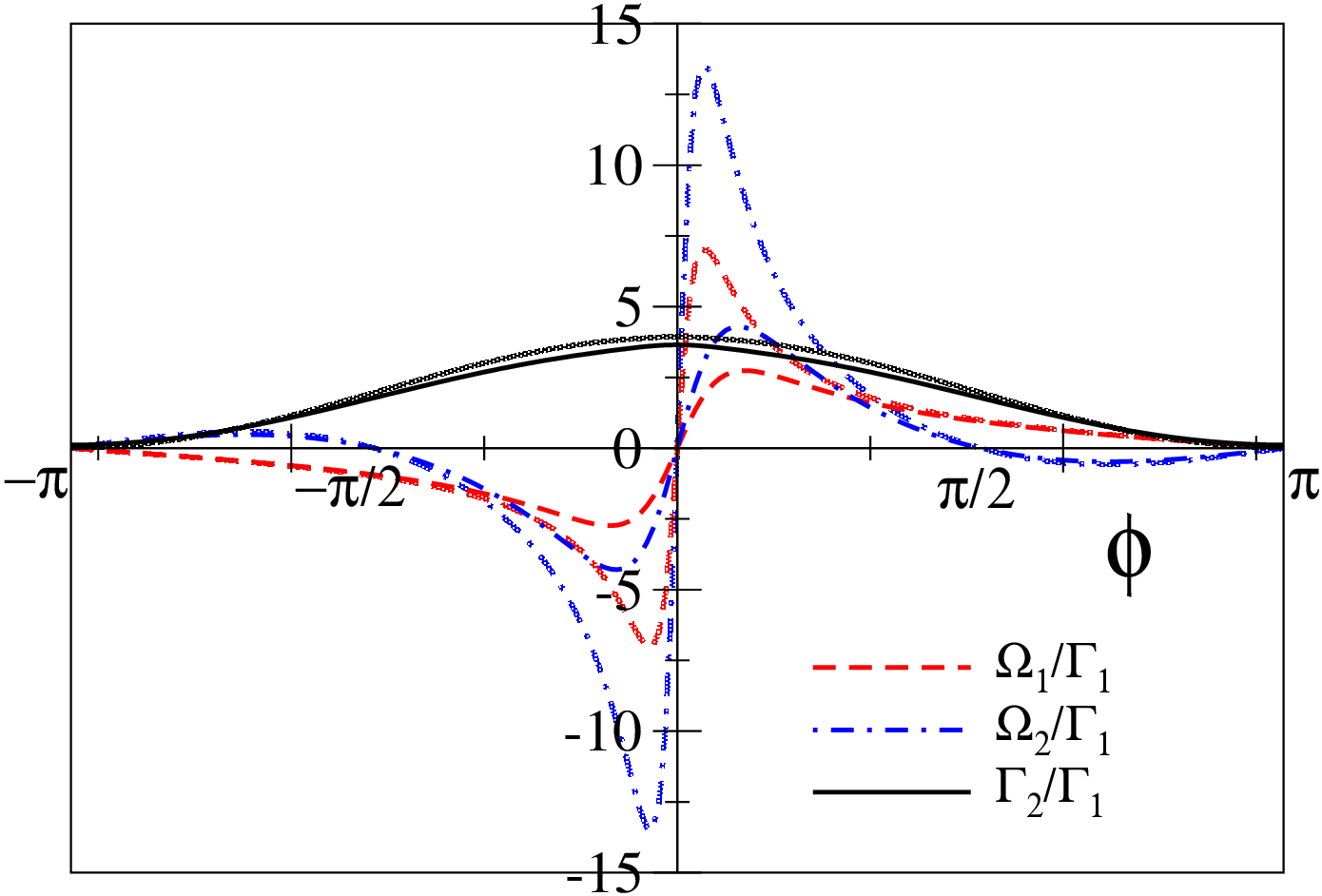}
 \caption{(Color online) The ratios $\Gamma_{2}/\Gamma_{1}$ {(solid line)}, $\Omega_{1}/\Gamma_{1}$ {(dashed line)} and $\Omega_{2}/\Gamma_{1}$ {(dashed-dotted 
line)} are shown as functions of the mode $\phi$, when 
$\widetilde{Q}(\varphi)$ is given by the wrapped-Gaussian distribution (wrapped stable distribution with index $\alpha=2$), with values of the scale parameter 
$\sigma=0.25$ ({thick-dark} lines) and 0.1 ({thin-}fuzzy lines).}
\label{Ratios_WrappedGaussian}
\end{figure}

The specific physical processes underlying the stationary scattering-angle distribution $\widetilde{Q}(\varphi)$, define the mode $\phi$.
We consider the effects of $\phi$ on the values of $\Gamma_{1}$, $\Gamma_{2}$, $\Omega_{1}$ and $\Omega_{2}$, (whose variations are not independent among 
them). For the particular case of the wrapped Gaussian ($\alpha=2$) with fixed scale parameter $\sigma=1/4$ and 1/10, and zero skewness, the ratios 
$\Gamma_{2}/\Gamma_{1}$, $\Omega_{1}/\Gamma_{1}$ and $\Omega_{2}/\Gamma_{1}$ 
are shown in Fig. \ref{Ratios_WrappedGaussian} (dark-thick lines correspond to $\sigma=0.25$, fuzzy-thin lines to $\sigma=0.1$) to be nonmonotonous functions 
of $\phi$.

The ratio 
$\Gamma_{2}/\Gamma_{1}$ (thick-solid-black line) is symmetric about $\phi=0$ and reaches its maximum and minimum values [see Eqs. \eqref{maxG2} and \eqref{minG2} in the 
appendix] at $\phi=0$ and $\phi=\pm\pi$ respectively.  The ratio $\Omega_{1}/\Gamma_{1}$ (thick-dashed-red line), which is antisymmetric about $\phi=0$ and gives the 
frequency of circular motion in units of $\Gamma_{1}$ induced by the distribution asymmetry, has a unique maximum value [given by Eq. \eqref{maxOmega1} in the appendix] at 
the mode $\phi=\arccos e^{-\sigma^{\alpha}}.$ This mode departs rapidly from the origin as $\sigma$ gets larger (for any $0<\alpha\le2$), saturating asymptotically at the 
value $\pi/2$. Thus, the larger frequency of active-circular motion is found at modes for which the scattered-angles is less than $\pi/2$. The ratio 
$\Omega_{2}/\Gamma_{1}$ (thick-dashed-dotted-blue line), is also antisymmetric about 
$\phi=0$ and exhibits a maximum and a minimum in the interval $[0,\pi];$ this occurs due to the two branches of the function $\cos2\phi/\cos^{3}\phi$ in Eq. 
\eqref{ExtremaOmega2} (see appendix) that determines the extrema values of the ratio $\Omega_{2}/\Gamma_{1}$.

\begin{figure}
 \includegraphics[width=\columnwidth]{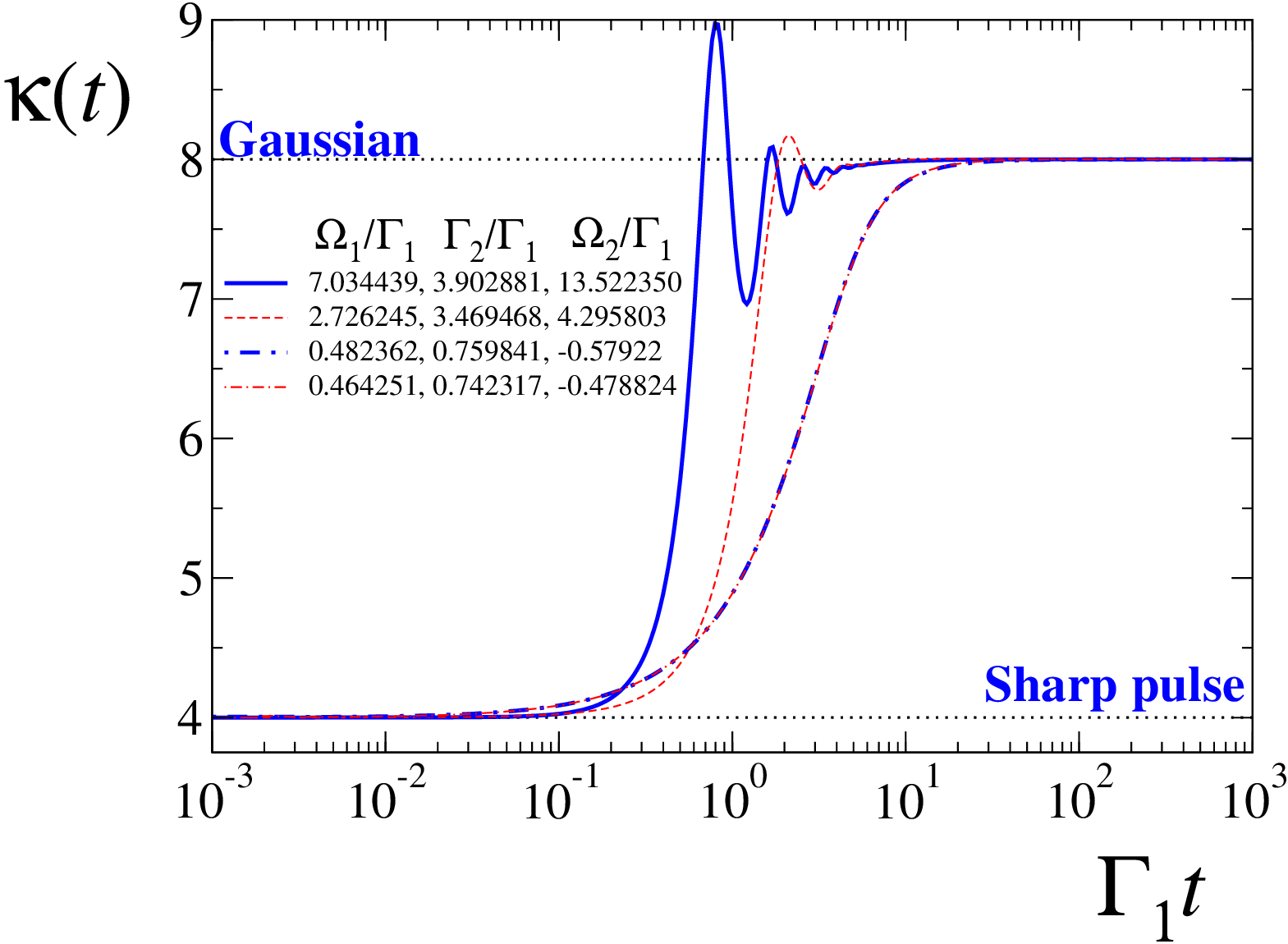}
 \caption{(Color online) The time dependence of the kurtosis is shown for $\widetilde{Q}(\varphi)$ given by the wrapped-Gaussian distribution (wrapped stable 
distribution with index $\alpha=2$), with values of the scale parameter $\sigma=0.1$ (thick-blue lines) and $0.25$ (thin-red lines). The values of the ratios: 
$\Gamma_{2}/\Gamma_{1}$, $\Omega_{1}/\Gamma_{1}$ and $\Omega_{2}/\Gamma_{1}$, correspond to those values of $\phi\in[0,\pi]$, for which $\Omega_{2}/\Gamma_{1}$ is maximum 
(solid-blue and dashed-red lines) and when is minimum (thick-dotted-dashed and thin-dotted-dashed lines), as can be noticed in Fig. \ref{Ratios_WrappedGaussian}.}
\label{KurtosisQAsym}
\end{figure}

As has been pointed out in previous sections, the kurtosis of the particle-position distribution carries information of the ratios  
$\Gamma_{2}/\Gamma_{1}$, $\Omega_{1}/\Gamma_{1}$ and $\Omega_{2}/\Gamma_{1}$, and, likewise, these ratios carry information about the asymmetry of the unimodal 
scattering-angle distribution \eqref{Qwrapped}, induced by the wrapped L\'evy $\alpha$-stable distributions \eqref{LevyDistributions}. In Fig.~\ref{KurtosisQAsym}, the time 
dependence of the kurtosis is shown for the values of the ratios obtained at the mode $\phi$, that makes  $\Omega_{2}/\Gamma_{1}$ to have its maximum and minimum value: 
$\phi_{\text{max}}\approx0.316$, $\phi_{\text{min}}\approx0.242$ for $\sigma=0.25$ (thin-red lines); and $\phi_{\text{max}}\approx0.139$, $\phi_{\text{min}}\approx2.238$ for 
$\sigma=0.1$ (thick-blue lines).

\subsection{Bimodal scattering-angle distributions}

It has been observed a variety of organisms that exhibit a bimodal distribution of scattering angles in their pattern of motion \cite{DuffyJBacter1997,ThevesBiophysJ2013}, 
and this bimodality has profound consequences on the spatial distributions of the particles. For the sake of clarifying this and in spite of the general analysis that can 
be carried out from our formalism for arbitrary scattered-angle distribution, we consider the limit case that corresponds to the 
bimodal distribution of scattered angles, with modes at the angles $\varphi_{1}$, $\varphi_{2}$, and of zero width, i.e.,
\begin{equation}
\widetilde{Q}(\varphi)=\nu\delta(\varphi-\varphi_{1})+(1-\nu)\delta(\varphi+\varphi_{2}), 
\end{equation}
where $0<\nu<1$ gives a weighing factor to each mode of the distribution. 

\subsubsection{Symmetrically distributed modes}
Consider the bimodal scattering-angle distribution of zero width 
\begin{equation}\label{bimodalSymm}
\widetilde{Q}(\varphi)=\nu\delta(\varphi-\varphi_{0})+(1-\nu)\delta(\varphi+\varphi_{0}), 
\end{equation}
where the modes are located symmetrically with respect to the forward direction at $\pm\varphi_{0}$ with $0<\varphi_{0}<\pi$; and $0<\nu<1$ gives the weight of each 
mode making the scattering-angle distribution asymmetric if $\nu\neq1/2$. It can be noticed from Eq. \eqref{Gamma-n} that 
$\Gamma_{n}$ is independent of $\nu$ for all $n$, having 
$\Lambda(1-\cos n\varphi_{0})$ as its value for given $n$ and $\varphi_{0}$. Also notice that the persistence time{, $\Gamma^{-1}$,} becomes arbitrarily large as 
$\varphi_{0}$ vanishes. In contrast, $\Omega_{n}$ does explicitly depend on $\nu$ as $\Lambda(2\nu-1)\sin n\varphi_{0}$.

\begin{figure}
 \includegraphics[width=\columnwidth]{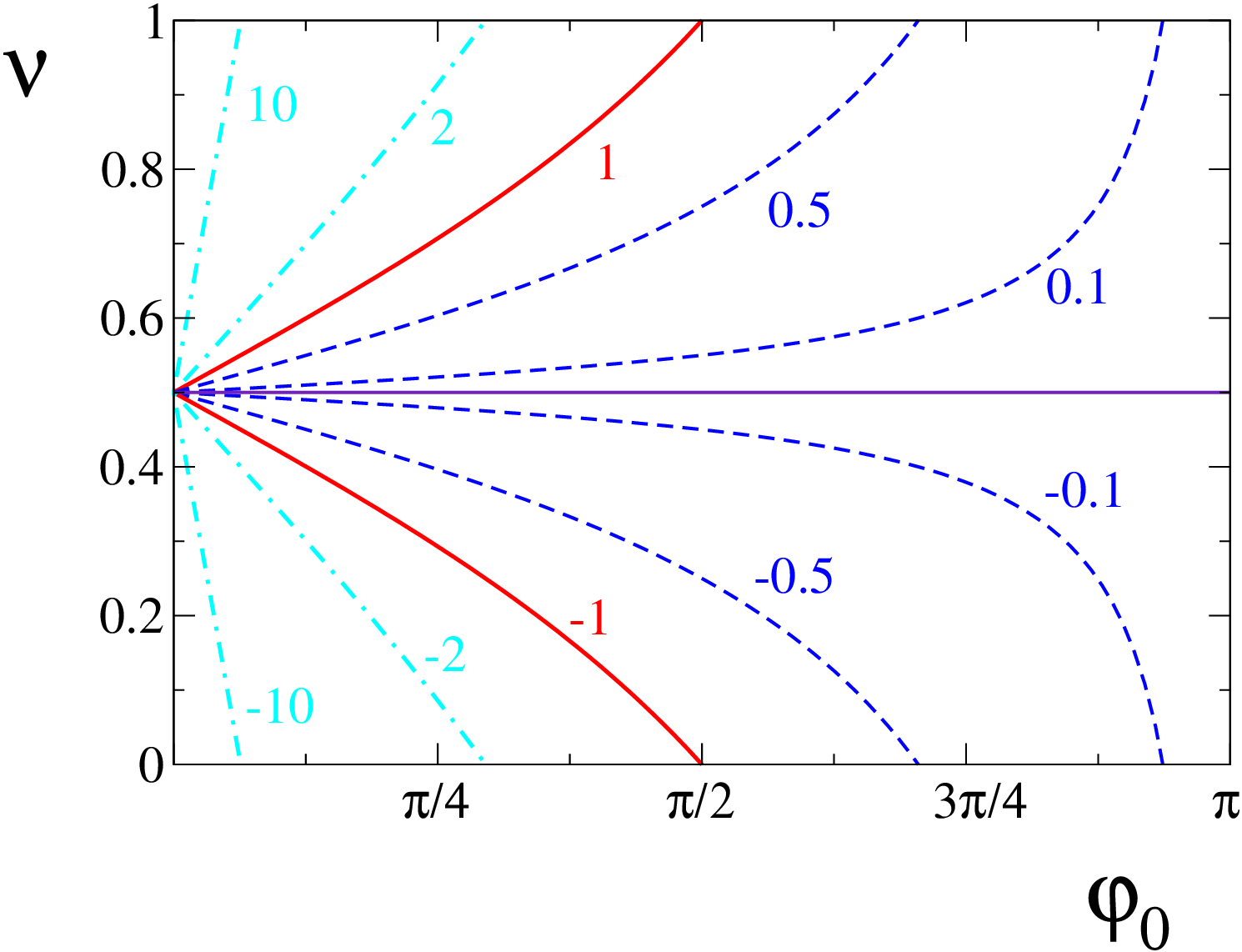}
 \caption{(Color online) Level curves of constant ratio $\chi=\Omega_{1}/\Gamma_{1}$ in the plane $\nu$-$\varphi_{0}$. The solid (red) line corresponds to the case 
$\Omega_{1}/\Gamma_{1}=1$ for which the effective diffusion coefficient is maximum. Dashed (blue) lines correspond to the cases for which 
$\vert\Omega_{1}/\Gamma_{1}\vert<1$ (shown $\chi=$10 and 2), while dashed-dotted (cyan) lines correspond to the cases for which $\vert\Omega_{1}/\Gamma_{1}\vert>1$ (shown 
$\chi=0.5$ and 0.1).}
\label{ContourPlot_Omega1Gamma1_Bimodal}
\end{figure}

The ratios $\Gamma_{2}/\Gamma_{1}$, $\Omega_{1}/\Gamma_{1}$ and $\Omega_{2}/\Gamma_{1}$, that give the full characterization of the kurtosis of the particle position 
distribution, can be calculated explicitly giving
\begin{subequations}
 \begin{align}
  \frac{\Gamma_{2}}{\Gamma_{1}}&=\left(2\cos\frac{\varphi_{0}}{2}\right)^{2},\\
  \frac{\Omega_{1}}{\Gamma_{1}}&=(2\nu-1)\cot\frac{\varphi_{0}}{2}\label{Omega1-SBimodal},\\
  \frac{\Omega_{2}}{\Gamma_{1}}&=2(2\nu-1)\cos\varphi_{0}\cot\frac{\varphi_{0}}{2}.
 \end{align}
\end{subequations}
From these expressions, several diffusive properties in terms of the parameters $\varphi_{0}$ and $\nu$ are obtained. First, after setting $\Omega_{1}/\Gamma_{1}=1$ in Eq. 
\eqref{Omega1-SBimodal}, the maximum value of the effective diffusion coefficient [see Eq. \eqref{DeffMax}] is obtained whenever 
$\nu=\left[1+\tan(\varphi_{0}/2)\right]/2$, with $0<\varphi_{0}<\pi/2$. The contour lines{, $\nu=\nu(\varphi_{0})$,} defined by fixing the ratio $\Omega_{1}/\Gamma_{1}$ 
to a constant $\chi$ are shown 
in Fig. \ref{ContourPlot_Omega1Gamma1_Bimodal}. 
\begin{figure}
 \includegraphics[width=\columnwidth]{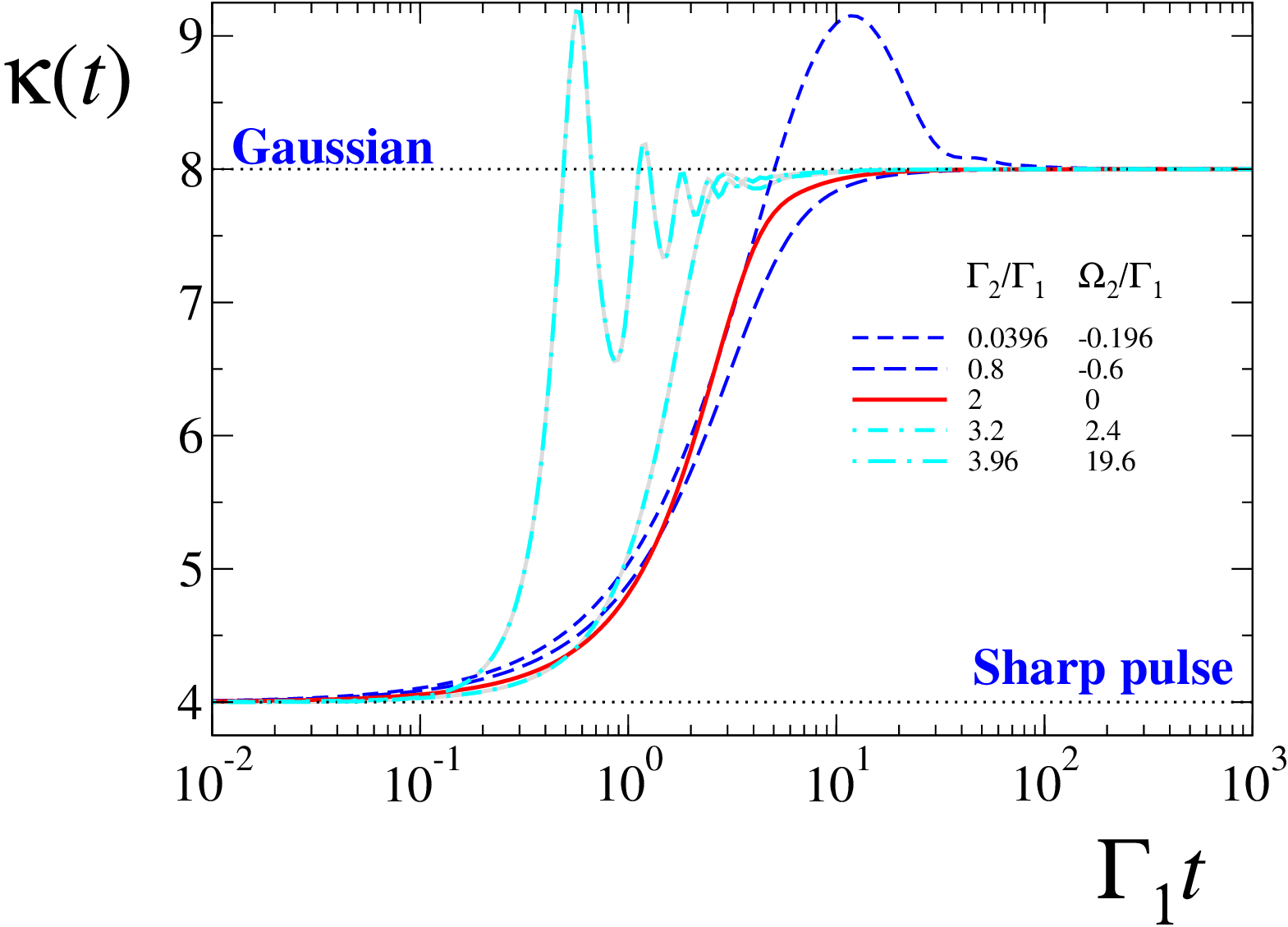}
 \caption{(Color online) Time dependence of the kurtosis $\kappa(t)$ for the bimodal distribution of scattered angles \eqref{bimodalSymm}. The values of ratios 
$\Gamma_{2}/\Gamma_{1}$, $\Omega_{2}/\Gamma_{1}$ correspond to the values of $\varphi_{0}^{*}$ that make $\nu=1$ 
on the contour lines given in Fig. \ref{ContourPlot_Omega1Gamma1_Bimodal} for the values of $\chi=10$, 2, 1, 0.5, and 0.1.}
\label{Kurtosis_vs_time_BimodalSymMoldes_maxOmega2}
\end{figure}

In Fig. \ref{Kurtosis_vs_time_BimodalSymMoldes_maxOmega2}, the time dependence of the kurtosis is shown as function of the dimensionless time $\Gamma_{1}t$, for the values 
of the ratios $\Gamma_{2}/\Gamma_{1}$ and $\Omega_{2}/\Gamma_{1}$, that correspond to the values of $\varphi_{0}^{*}$ that makes $\nu=1$ for a given ratio of 
$\chi=\Omega_{1}/\Gamma_{1}$: Oscillations are observed for times smaller or of the order of the persistence time for $\chi=10$, 2 (dashed-dotted lines), for 
which $\Gamma_{2}/\Gamma_{1}=3.96$, 3.2 and $\Omega_{2}/\Gamma_{1}=19.6$, 2.4 respectively. A smooth transition from a sharp pulse and the Gaussian distribution is observed 
for $\chi=1$ (maximum effective diffusion coefficient marked by the solid line), for which $\Gamma_{2}/\Gamma_{1}=2$ and $\Omega_{2}/\Gamma_{1}=0$. Such a transition is still 
smooth for $\chi=0.5$ (thick-dashed line, $\Gamma_{2}/\Gamma_{1}=0.8$, $\Omega_{2}/\Gamma_{1}=-0.6$). The  transition between the sharp pulse and the Gaussian distribution 
becomes nonmonotonic again, but now for times larger than the persistence time, when $\chi=0.1$ (thin-dashed line), for which $\Gamma_{2}/\Gamma_{1}=0.0396$ and 
$\Omega_{2}/\Gamma_{1}=-0.196$.

\subsubsection{Run-and-reverse}
Another instance of a simple bimodal distribution that can be analyzed to some detail, is given by the pattern of motion called \emph{run-and-reverse}. This pattern 
considers the scattering of the direction of motion along the forward and backward direction, thus having modes  at $\phi=0$ and $\pi$, respectively. In the case of zero 
width 
distribution, it can be written as
\begin{equation}\label{Qrun-and-reverse}
\widetilde{Q}(\varphi)=\nu\delta(\varphi)+(1-\nu)\delta(\varphi-\pi).
\end{equation}
Notice that $\Gamma_{n}$ vanishes for even $n$, and gives $2\Lambda (1-\nu)$ for all odd $n$, while $\Omega_{n}$ vanishes for all $n$. 
With this, the expansion \eqref{Pexpansion2} can be written as 
\begin{multline}\label{Pexpansion-run-and-reverse} 
\widetilde{P}(\boldsymbol{k},\varphi,t)=\frac{1}{2\pi}\widetilde{p}_{0}(\boldsymbol{k},t)+\frac{1}{2\pi}\sum_{n\, 
\text{even}}\widetilde{p}_{n}(\boldsymbol{k},t)e^{in\varphi}\\ 
+\frac{e^{ -\Gamma_ {1} t}}{2\pi}\sum_{n\, \text{odd}}\widetilde{p}_{n}(\boldsymbol{k},t)e^{in\varphi},
\end{multline}
from which a particular dynamics can be noticed, namely, there is a highly directional dependence in the long-time regime, as is evidenced by the fact that the second 
term in \eqref{Pexpansion-run-and-reverse} contributes to  $\widetilde{P}(\boldsymbol{k},\varphi,t)$ in such regime. This clearly contrasts with other patterns of active 
motion, for which $\widetilde{p}_{0}(\boldsymbol{k},t)$ gives the only contribution in the long-time regime as has been discussed in the previous sections. 
$\Gamma_{1}=2\Lambda(1-\nu),$ denotes the value of $\Gamma_{n}$, $n$ being odd.

\begin{figure}
 \includegraphics[width=\columnwidth]{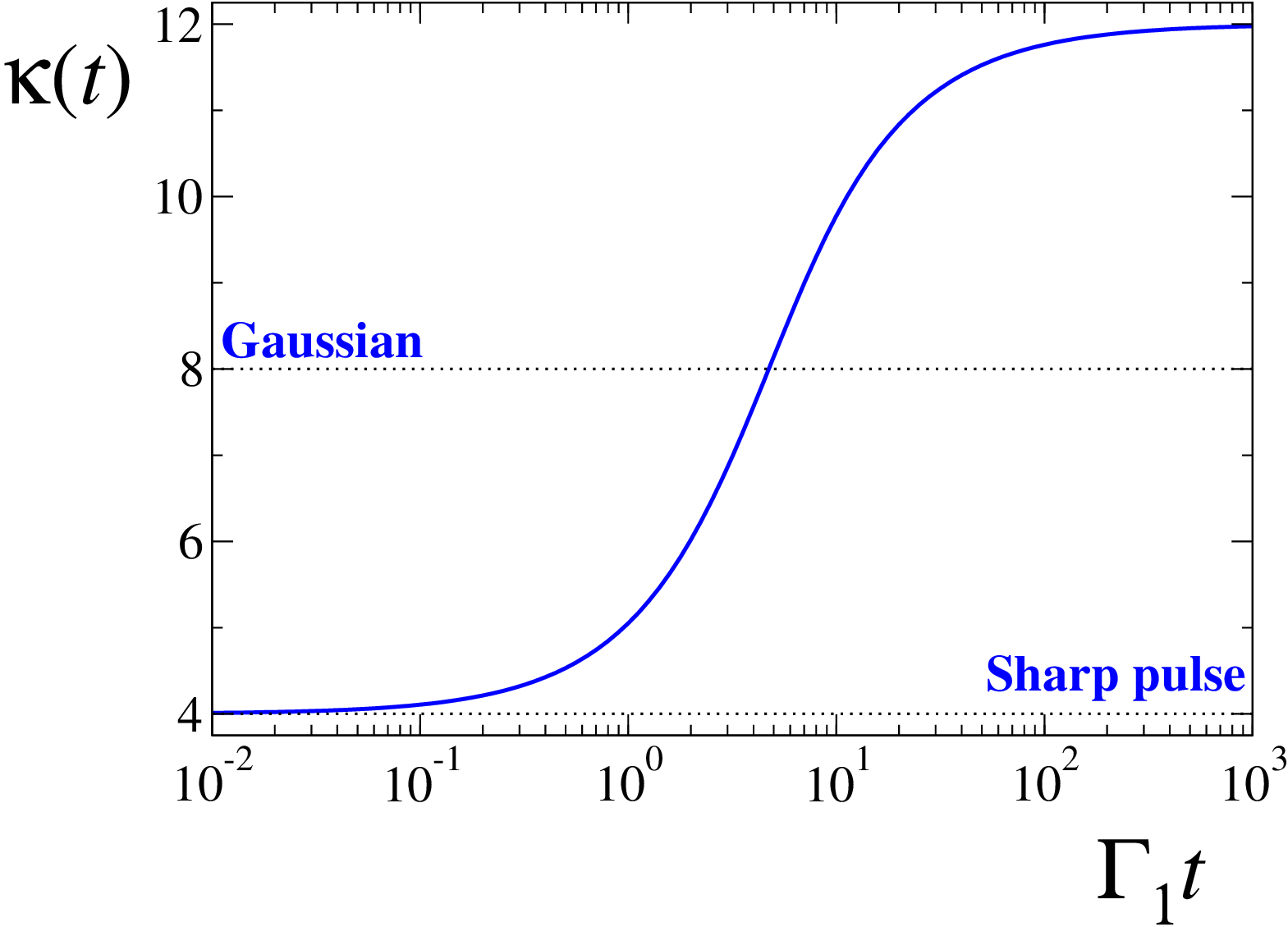}
 \caption{(Color online) The time dependence of the kurtosis is shown for the bimodal distribution with equally weighed modes at $0$ and $\pi$, which corresponds to a 
particular case of the patterns of motion called run-and-reverse.}
 \label{Kurtosis-Bimodal}
\end{figure}

The time dependence of the kurtosis can be obtained explicitly in this case,
\begin{equation}
 \kappa(t)=12\frac{6-4\Gamma_{1}  t+\Gamma_{1}^{2}t^{2}-2e^{-\Gamma_{1} t}(3+\Gamma_{1} t)}{[\Gamma_{1} t-(1-e^{-\Gamma_{1} t})]^{2}},
\end{equation}
and is shown in Fig. \ref{Kurtosis-Bimodal}. In the asymptotic limit{, the mean-squared displacement is linear in time with effective diffusion coefficient 
$v_{0}^{2}/4\Lambda(1-\nu)$, while} the kurtosis of the spatial distribution of the active particles goes to the value 12 (see Fig. \ref{Kurtosis-Bimodal}), which 
differs conspicuously from the value 8 that characterizes the two-dimensional Gaussian distribution. This scenario illustrates another instance of a diffusive 
process called ``anomalous, yet Brownian, diffusion'' 
\cite{WangPNAS2009,WangNatureMat2012,BhattacharyaJPCB2013}, which has been addressed theoretically in different one-dimensional models 
\cite{CressoniPRE2012,ChubynskyPRL2014,WangPRE2016} and in a three-dimensional study of chiral active motion \cite{SevillaPRE2016}.

For the scattering-angle distribution \eqref{Qrun-and-reverse} we have that the connecting function \eqref{Dke} 
acquires the simple form
\begin{equation}
\widetilde{\mathfrak{D}}(\boldsymbol{k},\epsilon)=\cfrac{2}{\epsilon+\Gamma_{\textcolor{red}{1}}
          + \cfrac{(v_{0}/2)^{2}\boldsymbol{k}^{2}}{\epsilon
          + \cfrac{(v_{0}/2)^{2}\boldsymbol{k}^{2}}{\epsilon+\Gamma_{\textcolor{red}{1}}
          +\cfrac{(v_{0}/2)^{2}\boldsymbol{k}^{2}}{\epsilon+\ddots
           }}}}
\end{equation}
which manifestly exhibits the singular role of the persistence time as the only timescale present in the dynamics and of the fact that in the long-time 
 regime ($\epsilon\rightarrow0$) no approximant of the connecting function is possible.
If the width at the modes of  Eq. \eqref{Qrun-and-reverse} is made finite, the  Gaussian distribution is recovered in the long-time regime since, the whole hierarchy of the
$\Gamma_{n}$ being recovered, the uniform scattering of the direction of motion in such a regime is assured.

\section{\label{SectConclusion}Conclusions}
I have presented a theoretical framework for the statistical analysis of the two-dimensional motion of active swimmers. This framework generalizes existent ones in that 
considers an arbitrary navigating strategy, that also takes into account circular motion, embedded in the arbitrary distribution of scattered angles of the particle's 
swimming direction. The framework is susceptible for generalizations, indeed, the transition rate of the direction of motion $K_{A}(\varphi\vert\varphi^{\prime})$ in 
Eq. \eqref{TransportEquation} can take into account a spatial and a temporal dependence, and also complements others that focus on the time distribution between fixed 
turning events. The method of solution presented, allowed for an exact analytical expression for the marginal probability distribution of finding a swimmer at $\boldsymbol{x}$ 
at time $t$, independently of the direction of motion. Such a solution can be cast as the exact solution of the generalized diffusion equation \eqref{Gral}, and an explicit 
expression for the time-space dependent memory function is presented. This result opens the door to consider the generalized diffusion equation \eqref{Gral} as a 
well-founded framework to analyze the motion of active swimmers.in particular, to consider time-space coupled memory function to describe other variety of patterns of active 
motion, as the ones described by L\'evy walks .

I also presented exact calculations for the time dependence of the mean-square displacement, which depends only on the ratio of the frequency of the circular motion induced 
by the specific scattering-angle distribution that embeds the pattern of active motion, $\Omega_{1}$, to the persistence time $\Gamma_{1}$. Certainly, there 
are plenty of patterns of motion that lead to the same ratio $\Omega_{1}/\Gamma_{1}$, and as such, the mean-square displacement is typical of many 
of them. However, the differences among different patterns of motion are unveiled in the \emph{intermediate-time regime} if more information of the pattern of motion is 
considered (as analyzed experimentally and theoretically for active Brownian motion and run-and-tumble particles in Ref. \cite{KurzthalerPRL2018}), and not only those related 
to $\langle\cos \varphi\rangle_{\widetilde{Q}}$ and $\langle\sin \varphi\rangle_{\widetilde{Q}}$, as is the case 
for the mean-square displacement.

It was shown that consideration of $\Gamma_{2}$ and $\Omega_{2}$, besides $\Gamma_{1}$ and $\Omega_{1}$,  is enough to distinguish some features among 
different patterns of 
motion. Certainly, knowledge of these quantities allows the exact calculation of the time dependence of the kurtosis, which gives information about the ``shape'' of the 
particle's position distribution. Some patterns of motion induce a smooth transition with time, from the initial sharp pulse, to the Gaussian of the long-time regime. Others 
deviate from this behavior and transit, from the initial sharp pulse to the Gaussian distribution, in a rather complex way characterized by oscillations.

Finally, although exact solutions to the Fokker-Planck equation \eqref{TransportEquation3} are known {for the particular case of the 
uniform scattering-angle distribution $\widetilde{Q}(\varphi)=(2\pi)^{-1}$ \cite{MartensEPJE2012} (and for the three-dimensional active Brownian motion 
\cite{KurzthalerSciRep2016})}, the analysis presented in this paper provides a broad understanding of the influence of an arbitrary pattern of motion on the 
statistical properties of the active swimmers, and encourages the development of more general theoretical frameworks of active motion that allow the incorporation of more 
general conditions.
  
\begin{acknowledgments}
The author kindly acknowledges Juan Manuel P\'erez Pe\~na for his interest during the initial part of the research presented in this paper. This work was supported by 
UNAM-PAPIIT IN114717. 
\end{acknowledgments}

\appendix

\section*{\label{Appendix}Appendix}
\subsection{The convolved solution \eqref{SolutionSeparation}}
Since the transport equation \eqref{TransportEquation} describes the diffusion process of an active particle in unconfined space, the Fourier transform can be applied to 
it, thus
\begin{multline}\label{TransportEquationFourier}
\frac{\partial }{\partial t}\widetilde{\mathcal{P}}({\boldsymbol{k}},\varphi ,t)+iv_{0}\hat{\boldsymbol{v}}\cdot 
\boldsymbol{k}\widetilde{\mathcal{P}}({\boldsymbol{k}},\varphi 
,t)=
-D_{T}\boldsymbol{k}^{2}\widetilde{\mathcal{P}}({\boldsymbol{k}},\varphi ,t)\\
+\int_{-\pi}^{\pi}d\varphi^{\prime} K_A\left(\varphi\vert\varphi^{\prime}\right)\widetilde{\mathcal{P}}(\boldsymbol{k},\varphi^{\prime},t).
\end{multline}
The solutions of the last equation admits the separation form
\begin{equation}\label{SolutionSeparationFourier}
\widetilde{\mathcal{P}}(\boldsymbol{k},\varphi,t)=\widetilde{G}_{D_{T}}(\boldsymbol{k},t)\widetilde{P}(\boldsymbol{k},\varphi,t),
\end{equation}
where $\widetilde{P}(\boldsymbol{k},\varphi,t)$ satisfies the equation
\begin{multline}\label{ActiveFPEFourier}
\frac{\partial }{\partial t}\widetilde{P}({\boldsymbol{k}},\varphi ,t)+iv_{0}\hat{\boldsymbol{v}}\cdot 
\boldsymbol{k}\widetilde{P}({\boldsymbol{k}},\varphi 
,t)=\\
\int_{-\pi}^{\pi}d\varphi^{\prime} K_A\left(\varphi\vert\varphi^{\prime}\right)\widetilde{P}(\boldsymbol{k},\varphi^{\prime},t),
\end{multline}
and $\widetilde{G}_{D_{T}}(\boldsymbol{k},t)=e^{-D_{T}\boldsymbol{k}^{2}t}$. Notice that expression \eqref{SolutionSeparation} corresponds to the inverse Fourier transform of 
the convolved solution \eqref{SolutionSeparationFourier}, Eq. \eqref{ActiveFPEFourier} corresponds to an equivalent form of the Fourier transform of Eq. \eqref{TransportEquation2}, and 
the Fourier inverse $\widetilde{G}_{D_{T}}(\boldsymbol{k},t)$, $G_{D_{T}}(\boldsymbol{x},t)$ satisfies the diffusion equation
\begin{equation}
\frac{\partial }{\partial t}G_{D_{T}}({\boldsymbol{x}},t)=D_{T}\nabla^{2}G_{D_{T}}(\boldsymbol{x},t).
\end{equation}

\subsection{Derivation of Eq. \eqref{p0}}
The explict and exact solution for $\widetilde{p}_{0}(\boldsymbol{k},\epsilon)$ given by Eq. \eqref{p0} is obtained as follows. After taking 
the Laplace transform of \eqref{Hierarchy} we have that this can be written as
\begin{multline}\label{RecurrenceRelation}
\widetilde{p}_{n}(\boldsymbol{k},\epsilon)+\frac{v_{0}}{2}ik\frac{1}{\epsilon}\Bigl[e^{-i\theta}\widetilde{p}_{n-1}(\boldsymbol{k},\epsilon+\lambda_{n-1}-\lambda_{n})\\
+e^{i\theta}\widetilde{p}_{n+1}(\boldsymbol{k},\epsilon+\lambda_{n+1}-\lambda_{n})\Bigr]=\frac{1}{\epsilon}\widetilde{p}_{n}^{(0)}(\boldsymbol{k})
\end{multline}
with $\widetilde{p}_{n}^{(0)}(\boldsymbol{k})=(2\pi)^{-1}\delta_{n,0}$ are the initial conditions. Notice that the corresponding Laplace argument of 
$\widetilde{p}_{n\pm1}(\boldsymbol{k},\epsilon)$ is shifted by $\lambda_{n\pm1}-\lambda_{n}$ respectively. 

For $n=0$ we have that
\begin{multline}\label{p0First}
\widetilde{p}_{0}(\boldsymbol{k},\epsilon)+\frac{v_{0}}{2}ik\frac{1}{\epsilon}\Bigl[e^{-i\theta}\widetilde{p}_{-1}(\boldsymbol{k},\epsilon+\lambda_{-1})\\
+e^{i\theta}\widetilde{p}_{1}(\boldsymbol{k},\epsilon+\lambda_{1})\Bigr]=\frac{1}{\epsilon}\widetilde{p}_{0}^{(0)}(\boldsymbol{k}),
\end{multline}
where we have used that $\lambda_{0}=0$. By use of the recurrence relation \eqref{RecurrenceRelation}, $\widetilde{p}_{\pm1}(\boldsymbol{k},\epsilon+\lambda_{\pm1})$ can be 
written in terms of $\widetilde{p}_{0}(\boldsymbol{k},\epsilon)$ and $\widetilde{p}_{\pm2}(\boldsymbol{k},\epsilon+\lambda_{\pm2})$ and the last equation can cast into
\begin{widetext}
\begin{multline}\label{p0Second}
\widetilde{p}_{0}(\boldsymbol{k},\epsilon)\Biggl[\epsilon+\Bigl(\frac{v_{0}}{2}\Bigr)^{2}\boldsymbol{k}^{2}\frac{1}{(\epsilon+\lambda_{1})}+\Bigl(\frac{v_{0}}{2}\Bigr)^{2}
\boldsymbol{k}^{2} \frac { 1 } { (\epsilon+\lambda_{-1})}\Biggr]+\\
\Bigl(\frac{v_{0}}{2}\Bigr)^{2}\boldsymbol{k}^{2}\Biggl[
\frac{e^ {-2i\theta}}{(\epsilon+\lambda_{-1})} \widetilde {p}_{ -2 } (\boldsymbol {k}, \epsilon+\lambda_ { -2 } )
+\frac{e^{2i\theta}}{(\epsilon+\lambda_{1})}\widetilde{p}_{2}(\boldsymbol{k},\epsilon+\lambda_{2})\Biggr]=\widetilde{p}_{0}^{(0)}(\boldsymbol{k}).
\end{multline}
In turn, $\widetilde{p}_{\pm2}(\boldsymbol{k},\epsilon+\lambda_{\pm2})$ can be written in terms of $\widetilde{p}_{0}(\boldsymbol{k},\epsilon)$ and 
$\widetilde{p}_{\pm3}(\boldsymbol{k},\epsilon+\lambda_{\pm3})$ by use of the recurrence relation \eqref{RecurrenceRelation}, and thus Eq. \eqref{p0Second} can turn into
\begin{multline}\label{p0Third}
\widetilde{p}_{0}(\boldsymbol{k},\epsilon)\left[\epsilon
          + \cfrac{(v_{0}/2)^{2}\boldsymbol{k}^{2}}{\epsilon+\lambda_{1}
          + \cfrac{(v_{0}/2)^{2}\boldsymbol{k}^{2}}{\epsilon+\lambda_{2}
           } }
	  +\cfrac{(v_{0}/2)^{2}\boldsymbol{k}^{2}}{\epsilon+\lambda_{-1}
          + \cfrac{(v_{0}/2)^{2}\boldsymbol{k}^{2}}{\epsilon+\lambda_{-2}
           } }\right]+\\
-i\Bigl(\frac{v_{0}}{2}\Bigr)^{3}\boldsymbol{k}^{2}k\Biggl[
\frac{e^ {-3i\theta}}{(\epsilon+\lambda_{-1})(\epsilon+\lambda_{-2})+(v_{0}/2)^{2}\boldsymbol{k}^{2}} \widetilde {p}_{ -3 } (\boldsymbol {k}, \epsilon+\lambda_ { -3 } )
\\
+\frac{e^{3i\theta}}{(\epsilon+\lambda_{1})(\epsilon+\lambda_{2})+(v_{0}/2)^{2}\boldsymbol{k}^{2}}\widetilde{p}_{3}(\boldsymbol{k},\epsilon+\lambda_{3})\Biggr]=\widetilde{p}_{
0 }^{(0)}(\boldsymbol{k}),
\end{multline}
\end{widetext}
and in turn, $\widetilde{p}_{\pm3}(\boldsymbol{k},\epsilon+\lambda_{\pm3})$ can be written in terms of $\widetilde{p}_{0}(\boldsymbol{k},\epsilon)$ and 
$\widetilde{p}_{\pm4}(\boldsymbol{k},\epsilon+\lambda_{\pm4})$ by use of the recurrence relation \eqref{RecurrenceRelation}, and so on. Thus the factor of 
$\widetilde{p}_{0}(\boldsymbol{k},\epsilon)$ corresponds to the denominator of Eq. \eqref{p0}.

\subsection{\label{appendix:MSD}The second and fourth moments of $p_{0}(\boldsymbol{x},t)$}
By use of the Eq. \eqref{MSDFormula}, the Laplace transform of the mean-squared displacement is obtained by substitution of the 
probability density $\widetilde{p}_{0}(\boldsymbol{k},\epsilon)$ given by Eq. \eqref{p0-2} with the initial condition $\widetilde{p}_{0}^{(0)}(k)=1/2\pi$, we have that
\begin{equation}
\langle\boldsymbol{x}^{2}(\epsilon)\rangle=-2\left.\frac{d^{2}}{dk^{2}}\left[\epsilon+(v_{0}/2)^{2}k^{2}\widetilde{\mathfrak{D}}(\boldsymbol{k},\epsilon)\right]^{
-1}\right\vert_{k=0}.
\end{equation}
After evaluation of the second-order derivative and evaluating at $k=0$ all terms proportional to $k$ and $k^{2}$ vanish, thus getting 
$\left[v_{0}^{2}\mathfrak{D}(,\epsilon)/\epsilon^{2}\right]_{k=0}$ which corresponds to Eq. \eqref{MSDLaplace}. With Eq. \eqref{Approximant1-FL} we get 
explicitly that
\begin{equation}
 \langle\boldsymbol{x}^{2}(\epsilon)\rangle=\frac{v_{0}^{2}}{\epsilon^{2}}\left[\frac{1}{\epsilon+\lambda_{1}}+\frac{1}{\epsilon+\lambda_{1}^{*}}\right],
\end{equation}
which can be inverted by the use of the convolution theorem of the Laplace transform, to have
\begin{equation}
 \langle\boldsymbol{x}^{2}(t)\rangle=v_{0}^{2}\int_{0}^{\infty}dt^{\prime}\left[e^{-\lambda_{1}(t-t^{\prime})}+e^{-\lambda_{1}^{*}(t-t^{\prime})}\right]t^{\prime}.
\end{equation}
By writing $\lambda_{1}=\Gamma_{1}+i\Omega_{1}$ and after evaluation of the elementary integrals we get Eq. \eqref{MSD}.

Analogously, the Laplace transform of the fourth moment is obtained from Eq. \eqref{RadialMoments} with $n=2$, i.e. from
\begin{equation}
  \langle x^{4}(\epsilon)\rangle _{\text{rad}}=\frac{8}{3}\left.\frac{d^{4}}{dk^{4}}\left[\epsilon+(v_{0}/2)^{2}k^{2}\widetilde{\mathfrak{D}}(\boldsymbol{k},\epsilon)\right]^{
-1}\right\vert_{k=0}.
\end{equation}
After evaluating the fourth-order derivative the only terms that do not vanish are given in Eq. \eqref{FourthMomentLaplace}. Finally, Eq. \eqref{2derivativeDke} is obtained 
as follows: From Eq. \eqref{Dke-Recursive}, we have that
\begin{equation}
\frac{\partial^{2}}{\partial k^{2}}\widetilde{\mathfrak{D}}(\boldsymbol{k},\epsilon)=\frac{\partial^{2}}{\partial 
k^{2}}\Delta_{0}(\boldsymbol{k},\epsilon)+\frac{\partial^{2}}{\partial k^{2}}\overline{\Delta}_{0}(\boldsymbol{k},\epsilon).
\end{equation}
From the definitions \eqref{RecursiveRels}, $\Delta_{0}(\boldsymbol{k},\epsilon)$ [and analogously $\overline{\Delta}_{0}(\boldsymbol{k},\epsilon)$] can be written as
\begin{equation}
 \Delta_{0}(\boldsymbol{k},\epsilon)=\frac{1}{\epsilon+\lambda_{1}+(v_{0}/2)^{2}k^{2}\Delta_{1}(\boldsymbol{k},\epsilon)}
\end{equation}
and thus, 
\begin{widetext}
      \begin{equation}
 \frac{\partial^{2}}{\partial 
k^{2}}\Delta_{0}(\boldsymbol{k},\epsilon) = 2(v_{0}/2)^{4}k^{2}
\frac{\left[2\Delta_{1}(\boldsymbol{k},\epsilon)+k\partial_{k}\Delta_{1}(\boldsymbol{k},\epsilon)\right]^{2}}{\left[\epsilon+\lambda_{1}+(v_{0}/2)^{2}k^{2}
\Delta_{ 1 } (\boldsymbol {k},\epsilon)\right]^{3}}-(v_{0}/2)^{2}
\frac{\left[2\Delta_{1}(\boldsymbol{k},\epsilon)+2k\partial_{k}\Delta_{1}(\boldsymbol{k},\epsilon)+k^{2}\partial_{k}^{2}\Delta_{1}(\boldsymbol{k},\epsilon)\right]}{\left[
\epsilon+\lambda_{1}+(v_{0}/2)^{2}k^{2}
\Delta_{ 1 } (\boldsymbol {k},\epsilon)\right]^{2}},
\end{equation}
     \end{widetext}
where $\partial_{k}$ denotes the partial derivative with respect $k$. A similar expression is obtained for $\left(\partial^{2}/\partial 
k^{2}\right)\overline{\Delta}_{0}(\boldsymbol{k},\epsilon)$. Thus, by evaluating these results at $k=0$, the expression
\begin{multline}
\left.\frac{\partial^{2}}{\partial 
k^{2}}\widetilde{\mathfrak{D}}(\boldsymbol{k},\epsilon)\right\vert_{k=0}=-\frac{v_{0}^{2}}{2}\frac{\Delta_{1}(\boldsymbol{k},\epsilon)\vert_{k=0}}{
(\epsilon+\lambda_{1})^{2}}\\
-\frac{v_{0}^{2}}{2}\frac{\overline{\Delta}_{1}(\boldsymbol{k},\epsilon)\vert_{k=0}}{
(\epsilon+\lambda_{1}^{*})^{2}}
\end{multline}
is obtained. From \eqref{RecursiveRels} we have that $\Delta_{1}(\boldsymbol{k},\epsilon)\vert_{k=0}=1/(\epsilon+\lambda_{2})$, and
$\overline{\Delta}_{1}(\boldsymbol{k},\epsilon)\vert_{k=0}=1/(\epsilon+\lambda_{2}^{*})$, and the Eq. \eqref{2derivativeDke} follows.

\subsection{\label{RatiosAppendix}Extrema of the ratios $\Gamma_{2}/\Gamma_{1}$, $\Omega_{1}/\Gamma_{1}$ and $\Omega_{2}/\Gamma_{1}$ for the L\'evy alpha-stable 
distributions}

From Eqs. \eqref{ScalesStableDist} we have the ratios 
\begin{subequations}\label{Ratios}
 \begin{align}
    \frac{\Gamma_{2}}{\Gamma_{1}}&=\frac{1-e^{-(2\sigma)^{\alpha}}\cos2\phi}{1-e^{-\sigma^{\alpha}}\cos\phi},\\
    \frac{\Omega_{1}}{\Gamma_{1}}&=\frac{e^{-\sigma^{\alpha}}\sin\phi}{1-e^{-\sigma^{\alpha}}\cos\phi},\\
    \frac{\Omega_{2}}{\Gamma_{1}}&=\frac{e^{-(2\sigma)^{\alpha}}\sin2\phi}{1-e^{-\sigma^{\alpha}}\cos\phi},
 \end{align}
\end{subequations}
in function of the mode $\phi$ for the distribution L\'evy alpha-stable distribution \eqref{LevyDistributions} with vanishing skewness [$\beta=0$ in Eq. 
\eqref{LevyDistributions}]. $\Gamma^{-1}$ gives account of the persistence time. The ratios \eqref{Ratios} are nonmotonous functions of $\phi$ and their 
corresponding extrema values are obtained from the equations
\begin{subequations}\label{ExtremaRatios}
 \begin{align}
    %&\sin\phi\left[4e^{-(2^{\alpha}-1)\sigma^{\alpha}}\cos\phi-\frac{1-e^{-(2\sigma)^{\alpha}}\cos2\phi}{1-e^{\sigma^{\alpha}}\cos\phi}\right]=0,\\
    &\sin\phi=0,\\
    &\cos\phi=e^{-\sigma^{\alpha}},\\
    &\frac{\cos2\phi}{\cos^{3}\phi}=e^{-\sigma^{\alpha}},\label{ExtremaOmega2}
 \end{align}
\end{subequations}
respectively.

The maximum value of $\Gamma_{2}/\Gamma_{1}$, 
\begin{equation}\label{maxG2}
\frac{1-e^{-(2\sigma)^{\alpha}}}{1-e^{-\sigma^{\alpha}}},
\end{equation}
occurs at $\phi=0$, while the minimum,
\begin{equation}\label{minG2}
\frac{1-e^{-(2\sigma)^{\alpha}}}{1+e^{-\sigma^{\alpha}}},
\end{equation}
occurs at $\phi=\pm\pi$ as is shown in Fig. \ref{Ratios_WrappedGaussian} for $\alpha=2$ and $\sigma=1/4,$ 1/10.

The ratio $\Omega_{1}/\Gamma_{1}$ has its maximum value
\begin{equation}\label{maxOmega1}
 \frac{e^{-\sigma^{\alpha}}\sin\arccos e^{-\sigma^{\alpha}}}{1-e^{-2\sigma^{\alpha}}}
\end{equation}
at the sole extreme $\phi=\arccos e^{-\sigma^{\alpha}}$.

% \bibliography{/home/francisco/Documents/MyBackUp_francisco/MyWork/biblio}

\begin{thebibliography}{49}
\expandafter\ifx\csname natexlab\endcsname\relax\def\natexlab#1{#1}\fi
\expandafter\ifx\csname bibnamefont\endcsname\relax
  \def\bibnamefont#1{#1}\fi
\expandafter\ifx\csname bibfnamefont\endcsname\relax
  \def\bibfnamefont#1{#1}\fi
\expandafter\ifx\csname citenamefont\endcsname\relax
  \def\citenamefont#1{#1}\fi
\expandafter\ifx\csname url\endcsname\relax
  \def\url#1{\texttt{#1}}\fi
\expandafter\ifx\csname urlprefix\endcsname\relax\def\urlprefix{URL }\fi
\providecommand{\bibinfo}[2]{#2}
\providecommand{\eprint}[2][]{\url{#2}}

\bibitem[{\citenamefont{Taktikos et~al.}(2014)\citenamefont{Taktikos, Stark,
  and Zaburdaev}}]{TaktikosPlos2014}
\bibinfo{author}{\bibfnamefont{J.}~\bibnamefont{Taktikos}},
  \bibinfo{author}{\bibfnamefont{H.}~\bibnamefont{Stark}}, \bibnamefont{and}
  \bibinfo{author}{\bibfnamefont{V.}~\bibnamefont{Zaburdaev}},
  \bibinfo{journal}{PLoS ONE} \textbf{\bibinfo{volume}{8}}, \bibinfo{pages}{1}
  (\bibinfo{year}{2014}),
  \urlprefix\url{http://dx.doi.org/10.1371%2Fjournal.pone.0081936}.

\bibitem[{\citenamefont{Detcheverry}(2015)}]{DetcheverryEPL2015}
\bibinfo{author}{\bibfnamefont{F.}~\bibnamefont{Detcheverry}},
  \bibinfo{journal}{EPL (Europhysics Letters)} \textbf{\bibinfo{volume}{111}},
  \bibinfo{pages}{60002} (\bibinfo{year}{2015}),
  \urlprefix\url{http://stacks.iop.org/0295-5075/111/i=6/a=60002}.

\bibitem[{\citenamefont{Bechinger et~al.}(2016)\citenamefont{Bechinger,
  Di~Leonardo, L\"owen, Reichhardt, Volpe, and Volpe}}]{BechingerRMP2016}
\bibinfo{author}{\bibfnamefont{C.}~\bibnamefont{Bechinger}},
  \bibinfo{author}{\bibfnamefont{R.}~\bibnamefont{Di~Leonardo}},
  \bibinfo{author}{\bibfnamefont{H.}~\bibnamefont{L\"owen}},
  \bibinfo{author}{\bibfnamefont{C.}~\bibnamefont{Reichhardt}},
  \bibinfo{author}{\bibfnamefont{G.}~\bibnamefont{Volpe}}, \bibnamefont{and}
  \bibinfo{author}{\bibfnamefont{G.}~\bibnamefont{Volpe}},
  \bibinfo{journal}{Rev. Mod. Phys.} \textbf{\bibinfo{volume}{88}},
  \bibinfo{pages}{045006} (\bibinfo{year}{2016}),
  \urlprefix\url{https://link.aps.org/doi/10.1103/RevModPhys.88.045006}.

\bibitem[{\citenamefont{Pototsky and Stark}(2012)}]{PototskyEPL2012}
\bibinfo{author}{\bibfnamefont{A.}~\bibnamefont{Pototsky}} \bibnamefont{and}
  \bibinfo{author}{\bibfnamefont{H.}~\bibnamefont{Stark}},
  \bibinfo{journal}{{EPL} (Europhysics Letters)} \textbf{\bibinfo{volume}{98}},
  \bibinfo{pages}{50004} (\bibinfo{year}{2012}),
  \urlprefix\url{https://doi.org/10.1209%2F0295-5075%2F98%2F50004}.

\bibitem[{\citenamefont{Cates and Tailleur}(2013)}]{CatesEPL2013}
\bibinfo{author}{\bibfnamefont{M.~E.} \bibnamefont{Cates}} \bibnamefont{and}
  \bibinfo{author}{\bibfnamefont{J.}~\bibnamefont{Tailleur}},
  \bibinfo{journal}{EPL (Europhysics Letters)} \textbf{\bibinfo{volume}{101}},
  \bibinfo{pages}{20010} (\bibinfo{year}{2013}),
  \urlprefix\url{http://stacks.iop.org/0295-5075/101/i=2/a=20010}.

\bibitem[{\citenamefont{Solon et~al.}(2015)\citenamefont{Solon, Cates, and
  Tailleur}}]{Solon2015EPJST2015}
\bibinfo{author}{\bibfnamefont{A.~P.} \bibnamefont{Solon}},
  \bibinfo{author}{\bibfnamefont{M.~E.} \bibnamefont{Cates}}, \bibnamefont{and}
  \bibinfo{author}{\bibfnamefont{J.}~\bibnamefont{Tailleur}},
  \bibinfo{journal}{The European Physical Journal Special Topics}
  \textbf{\bibinfo{volume}{224}}, \bibinfo{pages}{1231} (\bibinfo{year}{2015}),
  ISSN \bibinfo{issn}{1951-6401},
  \urlprefix\url{http://dx.doi.org/10.1140/epjst/e2015-02457-0}.

\bibitem[{\citenamefont{Stark}(2016)}]{StarkEPJST2016}
\bibinfo{author}{\bibfnamefont{H.}~\bibnamefont{Stark}}, \bibinfo{journal}{The
  European Physical Journal Special Topics} \textbf{\bibinfo{volume}{225}},
  \bibinfo{pages}{2369} (\bibinfo{year}{2016}), ISSN \bibinfo{issn}{1951-6401},
  \urlprefix\url{http://dx.doi.org/10.1140/epjst/e2016-60060-2}.
  
\bibitem[{\citenamefont{{Caprini Lorenzo} et~al.}(2019)\citenamefont{{Caprini
  Lorenzo}, {Marini Bettolo Marconi Umberto}, and {Puglisi
  Andrea}}}]{CapriniSciRep2019}
\bibinfo{author}{\bibnamefont{{Caprini Lorenzo}}},
  \bibinfo{author}{\bibnamefont{{Marini Bettolo Marconi Umberto}}},
  \bibnamefont{and} \bibinfo{author}{\bibnamefont{{Puglisi Andrea}}},
  \bibinfo{journal}{Scientific Reports} \textbf{\bibinfo{volume}{9}},
  \bibinfo{pages}{1386} (\bibinfo{year}{2019}), ISSN \bibinfo{issn}{2045-2322}.
  
\bibitem[{\citenamefont{Wu and Libchaber}(2000)}]{WuPRL2000}
\bibinfo{author}{\bibfnamefont{X.-L.} \bibnamefont{Wu}} \bibnamefont{and}
  \bibinfo{author}{\bibfnamefont{A.}~\bibnamefont{Libchaber}},
  \bibinfo{journal}{Phys. Rev. Lett.} \textbf{\bibinfo{volume}{84}},
  \bibinfo{pages}{3017} (\bibinfo{year}{2000}),
  \urlprefix\url{http://link.aps.org/doi/10.1103/PhysRevLett.84.3017}.

\bibitem[{\citenamefont{Szamel}(2014)}]{SzamelPRE2014}
\bibinfo{author}{\bibfnamefont{G.}~\bibnamefont{Szamel}},
  \bibinfo{journal}{Phys. Rev. E} \textbf{\bibinfo{volume}{90}},
  \bibinfo{pages}{012111} (\bibinfo{year}{2014}),
  \urlprefix\url{http://link.aps.org/doi/10.1103/PhysRevE.90.012111}.

\bibitem[{\citenamefont{Fodor et~al.}(2016)\citenamefont{Fodor, Nardini, Cates,
  Tailleur, Visco, and van Wijland}}]{FodorPRL2016}
\bibinfo{author}{\bibfnamefont{E.}~\bibnamefont{Fodor}},
  \bibinfo{author}{\bibfnamefont{C.}~\bibnamefont{Nardini}},
  \bibinfo{author}{\bibfnamefont{M.~E.} \bibnamefont{Cates}},
  \bibinfo{author}{\bibfnamefont{J.}~\bibnamefont{Tailleur}},
  \bibinfo{author}{\bibfnamefont{P.}~\bibnamefont{Visco}}, \bibnamefont{and}
  \bibinfo{author}{\bibfnamefont{F.}~\bibnamefont{van Wijland}},
  \bibinfo{journal}{Phys. Rev. Lett.} \textbf{\bibinfo{volume}{117}},
  \bibinfo{pages}{038103} (\bibinfo{year}{2016}),
  \urlprefix\url{https://link.aps.org/doi/10.1103/PhysRevLett.117.038103}.

\bibitem[{\citenamefont{{Maggi Claudio} et~al.}(2015)\citenamefont{{Maggi
  Claudio}, {Marconi Umberto Marini Bettolo}, {Gnan Nicoletta}, and {Di
  Leonardo Roberto}}}]{MaggiSciRep2015}
\bibinfo{author}{\bibnamefont{{Maggi Claudio}}},
  \bibinfo{author}{\bibnamefont{{Marconi Umberto Marini Bettolo}}},
  \bibinfo{author}{\bibnamefont{{Gnan Nicoletta}}}, \bibnamefont{and}
  \bibinfo{author}{\bibnamefont{{Di Leonardo Roberto}}},
  \bibinfo{journal}{Scientific Reports} \textbf{\bibinfo{volume}{5}},
  \bibinfo{pages}{10742} (\bibinfo{year}{2015}), ISSN
  \bibinfo{issn}{2045-2322}.

\bibitem[{\citenamefont{Farage et~al.}(2015)\citenamefont{Farage, Krinninger,
  and Brader}}]{FaragePRE2015}
\bibinfo{author}{\bibfnamefont{T.~F.~F.} \bibnamefont{Farage}},
  \bibinfo{author}{\bibfnamefont{P.}~\bibnamefont{Krinninger}},
  \bibnamefont{and} \bibinfo{author}{\bibfnamefont{J.~M.}
  \bibnamefont{Brader}}, \bibinfo{journal}{Phys. Rev. E}
  \textbf{\bibinfo{volume}{91}}, \bibinfo{pages}{042310}
  (\bibinfo{year}{2015}),
  \urlprefix\url{https://link.aps.org/doi/10.1103/PhysRevE.91.042310}.

\bibitem[{\citenamefont{Paoluzzi et~al.}(2016)\citenamefont{Paoluzzi, Maggi,
  Marini Bettolo~Marconi, and Gnan}}]{PaoluzziPRE2016}
\bibinfo{author}{\bibfnamefont{M.}~\bibnamefont{Paoluzzi}},
  \bibinfo{author}{\bibfnamefont{C.}~\bibnamefont{Maggi}},
  \bibinfo{author}{\bibfnamefont{U.}~\bibnamefont{Marini Bettolo~Marconi}},
  \bibnamefont{and} \bibinfo{author}{\bibfnamefont{N.}~\bibnamefont{Gnan}},
  \bibinfo{journal}{Phys. Rev. E} \textbf{\bibinfo{volume}{94}},
  \bibinfo{pages}{052602} (\bibinfo{year}{2016}),
  \urlprefix\url{https://link.aps.org/doi/10.1103/PhysRevE.94.052602}.

\bibitem[{\citenamefont{Sevilla and Castro-Villarreal}(2019)}]{Sevilla2019c}
\bibinfo{author}{\bibfnamefont{F.~J.} \bibnamefont{Sevilla}} \bibnamefont{and}
  \bibinfo{author}{\bibfnamefont{P.}~\bibnamefont{Castro-Villarreal}},
  \bibinfo{journal}{arXiv preprint arXiv:1912.03425}  (\bibinfo{year}{2019}).

\bibitem[{\citenamefont{Kurzthaler et~al.}(2018)\citenamefont{Kurzthaler,
  Devailly, Arlt, Franosch, Poon, Martinez, and Brown}}]{KurzthalerPRL2018}
\bibinfo{author}{\bibfnamefont{C.}~\bibnamefont{Kurzthaler}},
  \bibinfo{author}{\bibfnamefont{C.}~\bibnamefont{Devailly}},
  \bibinfo{author}{\bibfnamefont{J.}~\bibnamefont{Arlt}},
  \bibinfo{author}{\bibfnamefont{T.}~\bibnamefont{Franosch}},
  \bibinfo{author}{\bibfnamefont{W.~C.~K.} \bibnamefont{Poon}},
  \bibinfo{author}{\bibfnamefont{V.~A.} \bibnamefont{Martinez}},
  \bibnamefont{and} \bibinfo{author}{\bibfnamefont{A.~T.} \bibnamefont{Brown}},
  \bibinfo{journal}{Phys. Rev. Lett.} \textbf{\bibinfo{volume}{121}},
  \bibinfo{pages}{078001} (\bibinfo{year}{2018}),
  \urlprefix\url{https://link.aps.org/doi/10.1103/PhysRevLett.121.078001}.

 
\bibitem[{\citenamefont{Detcheverry}(2017)}]{DetcheverryPRE2017}
\bibinfo{author}{\bibfnamefont{F.~m.~c.} \bibnamefont{Detcheverry}},
  \bibinfo{journal}{Phys. Rev. E} \textbf{\bibinfo{volume}{96}},
  \bibinfo{pages}{012415} (\bibinfo{year}{2017}),
  \urlprefix\url{https://link.aps.org/doi/10.1103/PhysRevE.96.012415}.

\bibitem[{\citenamefont{Duderstadt and
  Martin}(1979)}]{DuderstadtTransportTheory}
\bibinfo{author}{\bibfnamefont{J.~J.} \bibnamefont{Duderstadt}}
  \bibnamefont{and} \bibinfo{author}{\bibfnamefont{W.~R.}
  \bibnamefont{Martin}}, \emph{\bibinfo{title}{{Transport theory.}}},
  vol.~\bibinfo{volume}{1} (\bibinfo{year}{1979}).

\bibitem[{\citenamefont{Porra et~al.}(1997)\citenamefont{Porra, Masoliver, and
  Weiss}}]{PorraPRE1997}
\bibinfo{author}{\bibfnamefont{J.~M.} \bibnamefont{Porra}},
  \bibinfo{author}{\bibfnamefont{J.}~\bibnamefont{Masoliver}},
  \bibnamefont{and} \bibinfo{author}{\bibfnamefont{G.~H.} \bibnamefont{Weiss}},
  \bibinfo{journal}{Physical Review E} \textbf{\bibinfo{volume}{55}},
  \bibinfo{pages}{7771} (\bibinfo{year}{1997}).

\bibitem[{\citenamefont{Figueroa-Morales
  et~al.}(2018)\citenamefont{Figueroa-Morales, Darnige, Douarche, Martinez,
  Soto, Lindner, and Cl{\'e}ment}}]{Figueroa2018}
\bibinfo{author}{\bibfnamefont{N.}~\bibnamefont{Figueroa-Morales}},
  \bibinfo{author}{\bibfnamefont{T.}~\bibnamefont{Darnige}},
  \bibinfo{author}{\bibfnamefont{C.}~\bibnamefont{Douarche}},
  \bibinfo{author}{\bibfnamefont{V.}~\bibnamefont{Martinez}},
  \bibinfo{author}{\bibfnamefont{R.}~\bibnamefont{Soto}},
  \bibinfo{author}{\bibfnamefont{A.}~\bibnamefont{Lindner}}, \bibnamefont{and}
  \bibinfo{author}{\bibfnamefont{E.}~\bibnamefont{Cl{\'e}ment}},
  \bibinfo{journal}{arXiv preprint arXiv:1803.01295}  (\bibinfo{year}{2018}).

  \bibitem[{\citenamefont{Sevilla et~al.}(2019)\citenamefont{Sevilla,
  Rodr\'{\i}guez, and Gomez-Solano}}]{SevillaPRE2019b}
\bibinfo{author}{\bibfnamefont{F.~J.} \bibnamefont{Sevilla}},
  \bibinfo{author}{\bibfnamefont{R.~F.} \bibnamefont{Rodr\'{\i}guez}},
  \bibnamefont{and} \bibinfo{author}{\bibfnamefont{J.~R.}
  \bibnamefont{Gomez-Solano}}, \bibinfo{journal}{Phys. Rev. E}
  \textbf{\bibinfo{volume}{100}}, \bibinfo{pages}{032123}
  (\bibinfo{year}{2019}),
  \urlprefix\url{https://link.aps.org/doi/10.1103/PhysRevE.100.032123}.

\bibitem[{\citenamefont{Zhang et~al.}(2010)\citenamefont{Zhang,
  Be{\textquoteright}er, Florin, and Swinney}}]{ZhangPNAS2010}
\bibinfo{author}{\bibfnamefont{H.~P.} \bibnamefont{Zhang}},
  \bibinfo{author}{\bibfnamefont{A.}~\bibnamefont{Be{\textquoteright}er}},
  \bibinfo{author}{\bibfnamefont{E.-L.} \bibnamefont{Florin}},
  \bibnamefont{and} \bibinfo{author}{\bibfnamefont{H.~L.}
  \bibnamefont{Swinney}}, \bibinfo{journal}{Proceedings of the National Academy
  of Sciences} \textbf{\bibinfo{volume}{107}}, \bibinfo{pages}{13626}
  (\bibinfo{year}{2010}), ISSN \bibinfo{issn}{0027-8424},
  \eprint{https://www.pnas.org/content/107/31/13626.full.pdf},
  \urlprefix\url{https://www.pnas.org/content/107/31/13626}.

\bibitem[{\citenamefont{Thutupalli et~al.}(2015)\citenamefont{Thutupalli, Sun,
  Bunyak, Palaniappan, and Shaevitz}}]{ThutupalliJRSI2015}
\bibinfo{author}{\bibfnamefont{S.}~\bibnamefont{Thutupalli}},
  \bibinfo{author}{\bibfnamefont{M.}~\bibnamefont{Sun}},
  \bibinfo{author}{\bibfnamefont{F.}~\bibnamefont{Bunyak}},
  \bibinfo{author}{\bibfnamefont{K.}~\bibnamefont{Palaniappan}},
  \bibnamefont{and} \bibinfo{author}{\bibfnamefont{J.~W.}
  \bibnamefont{Shaevitz}}, \bibinfo{journal}{Journal of The Royal Society
  Interface} \textbf{\bibinfo{volume}{12}}, \bibinfo{pages}{20150049}
  (\bibinfo{year}{2015}),
  \eprint{https://royalsocietypublishing.org/doi/pdf/10.1098/rsif.2015.0049},
  \urlprefix\url{https://royalsocietypublishing.org/doi/abs/10.1098/rsif.2015.0049}.

\bibitem[{\citenamefont{{Ginot F.} et~al.}(2018)\citenamefont{{Ginot F.},
  {Theurkauff I.}, {Detcheverry F.}, {Ybert C.}, and {Cottin-Bizonne
  C.}}}]{GinotNatureComm2018}
\bibinfo{author}{\bibnamefont{{Ginot F.}}},
  \bibinfo{author}{\bibnamefont{{Theurkauff I.}}},
  \bibinfo{author}{\bibnamefont{{Detcheverry F.}}},
  \bibinfo{author}{\bibnamefont{{Ybert C.}}}, \bibnamefont{and}
  \bibinfo{author}{\bibnamefont{{Cottin-Bizonne C.}}}, \bibinfo{journal}{Nature
  Communications} \textbf{\bibinfo{volume}{9}}, \bibinfo{pages}{696}
  (\bibinfo{year}{2018}), ISSN \bibinfo{issn}{2041-1723}.

\bibitem[{\citenamefont{Mahault et~al.}(2018)\citenamefont{Mahault, Jiang,
  Bertin, Ma, Patelli, Shi, and Chat\'e}}]{MahaultPRL2018}
\bibinfo{author}{\bibfnamefont{B.}~\bibnamefont{Mahault}},
  \bibinfo{author}{\bibfnamefont{X.-c.} \bibnamefont{Jiang}},
  \bibinfo{author}{\bibfnamefont{E.}~\bibnamefont{Bertin}},
  \bibinfo{author}{\bibfnamefont{Y.-q.} \bibnamefont{Ma}},
  \bibinfo{author}{\bibfnamefont{A.}~\bibnamefont{Patelli}},
  \bibinfo{author}{\bibfnamefont{X.-q.} \bibnamefont{Shi}}, \bibnamefont{and}
  \bibinfo{author}{\bibfnamefont{H.}~\bibnamefont{Chat\'e}},
  \bibinfo{journal}{Phys. Rev. Lett.} \textbf{\bibinfo{volume}{120}},
  \bibinfo{pages}{258002} (\bibinfo{year}{2018}),
  \urlprefix\url{https://link.aps.org/doi/10.1103/PhysRevLett.120.258002}.

\bibitem[{\citenamefont{Jeckel et~al.}(2019)\citenamefont{Jeckel, Jelli,
  Hartmann, Singh, Mok, Totz, Vidakovic, Eckhardt, Dunkel, and
  Drescher}}]{JeckelPNAS2018}
\bibinfo{author}{\bibfnamefont{H.}~\bibnamefont{Jeckel}},
  \bibinfo{author}{\bibfnamefont{E.}~\bibnamefont{Jelli}},
  \bibinfo{author}{\bibfnamefont{R.}~\bibnamefont{Hartmann}},
  \bibinfo{author}{\bibfnamefont{P.~K.} \bibnamefont{Singh}},
  \bibinfo{author}{\bibfnamefont{R.}~\bibnamefont{Mok}},
  \bibinfo{author}{\bibfnamefont{J.~F.} \bibnamefont{Totz}},
  \bibinfo{author}{\bibfnamefont{L.}~\bibnamefont{Vidakovic}},
  \bibinfo{author}{\bibfnamefont{B.}~\bibnamefont{Eckhardt}},
  \bibinfo{author}{\bibfnamefont{J.}~\bibnamefont{Dunkel}}, \bibnamefont{and}
  \bibinfo{author}{\bibfnamefont{K.}~\bibnamefont{Drescher}},
  \bibinfo{journal}{Proceedings of the National Academy of Sciences}
  \textbf{\bibinfo{volume}{116}}, \bibinfo{pages}{1489} (\bibinfo{year}{2019}),
  ISSN \bibinfo{issn}{0027-8424},
  \eprint{https://www.pnas.org/content/116/5/1489.full.pdf},
  \urlprefix\url{https://www.pnas.org/content/116/5/1489}.
  
\bibitem[{\citenamefont{Liu et~al.}(2019)\citenamefont{Liu, Patch, Bahar,
  Yllanes, Welch, Marchetti, Thutupalli, and Shaevitz}}]{LiuPRL2019}
\bibinfo{author}{\bibfnamefont{G.}~\bibnamefont{Liu}},
  \bibinfo{author}{\bibfnamefont{A.}~\bibnamefont{Patch}},
  \bibinfo{author}{\bibfnamefont{F.}~\bibnamefont{Bahar}},
  \bibinfo{author}{\bibfnamefont{D.}~\bibnamefont{Yllanes}},
  \bibinfo{author}{\bibfnamefont{R.~D.} \bibnamefont{Welch}},
  \bibinfo{author}{\bibfnamefont{M.~C.} \bibnamefont{Marchetti}},
  \bibinfo{author}{\bibfnamefont{S.}~\bibnamefont{Thutupalli}},
  \bibnamefont{and} \bibinfo{author}{\bibfnamefont{J.~W.}
  \bibnamefont{Shaevitz}}, \bibinfo{journal}{Phys. Rev. Lett.}
  \textbf{\bibinfo{volume}{122}}, \bibinfo{pages}{248102}
  (\bibinfo{year}{2019}),
  \urlprefix\url{https://link.aps.org/doi/10.1103/PhysRevLett.122.248102}.
  
\bibitem[{\citenamefont{Theves et~al.}(2013)\citenamefont{Theves, Taktikos,
  Zaburdaev, Stark, and Beta}}]{ThevesBiophysJ2013}
\bibinfo{author}{\bibfnamefont{M.}~\bibnamefont{Theves}},
  \bibinfo{author}{\bibfnamefont{J.}~\bibnamefont{Taktikos}},
  \bibinfo{author}{\bibfnamefont{V.}~\bibnamefont{Zaburdaev}},
  \bibinfo{author}{\bibfnamefont{H.}~\bibnamefont{Stark}}, \bibnamefont{and}
  \bibinfo{author}{\bibfnamefont{C.}~\bibnamefont{Beta}},
  \bibinfo{journal}{Biophysical Journal} \textbf{\bibinfo{volume}{105}},
  \bibinfo{pages}{1915 } (\bibinfo{year}{2013}), ISSN
  \bibinfo{issn}{0006-3495},
  \urlprefix\url{http://www.sciencedirect.com/science/article/pii/S0006349513010217}.

\bibitem[{\citenamefont{Viswanathan et~al.}(2005)\citenamefont{Viswanathan,
  Raposo, Bartumeus, Catalan, and da~Luz}}]{ViswanathanPRE2005}
\bibinfo{author}{\bibfnamefont{G.~M.} \bibnamefont{Viswanathan}},
  \bibinfo{author}{\bibfnamefont{E.~P.} \bibnamefont{Raposo}},
  \bibinfo{author}{\bibfnamefont{F.}~\bibnamefont{Bartumeus}},
  \bibinfo{author}{\bibfnamefont{J.}~\bibnamefont{Catalan}}, \bibnamefont{and}
  \bibinfo{author}{\bibfnamefont{M.~G.~E.} \bibnamefont{da~Luz}},
  \bibinfo{journal}{Phys. Rev. E} \textbf{\bibinfo{volume}{72}},
  \bibinfo{pages}{011111} (\bibinfo{year}{2005}),
  \urlprefix\url{http://link.aps.org/doi/10.1103/PhysRevE.72.011111}.

\bibitem[{\citenamefont{Bartumeus et~al.}(2008)\citenamefont{Bartumeus,
  Catalan, Viswanathan, Raposo, and da~Luz}}]{BartumeusJTB2008}
\bibinfo{author}{\bibfnamefont{F.}~\bibnamefont{Bartumeus}},
  \bibinfo{author}{\bibfnamefont{J.}~\bibnamefont{Catalan}},
  \bibinfo{author}{\bibfnamefont{G.}~\bibnamefont{Viswanathan}},
  \bibinfo{author}{\bibfnamefont{E.}~\bibnamefont{Raposo}}, \bibnamefont{and}
  \bibinfo{author}{\bibfnamefont{M.}~\bibnamefont{da~Luz}},
  \bibinfo{journal}{Journal of Theoretical Biology}
  \textbf{\bibinfo{volume}{252}}, \bibinfo{pages}{43 } (\bibinfo{year}{2008}),
  ISSN \bibinfo{issn}{0022-5193},
  \urlprefix\url{http://www.sciencedirect.com/science/article/pii/S0022519308000180}.

\bibitem[{\citenamefont{Sevilla and Nava}(2014)}]{SevillaPRE2014}
\bibinfo{author}{\bibfnamefont{F.~J.}~\bibnamefont{Sevilla}} \bibnamefont{and}
  \bibinfo{author}{\bibfnamefont{L.~A.} \bibnamefont{G\'omez~Nava}},
  \bibinfo{journal}{Physical Review E} \textbf{\bibinfo{volume}{90}},
  \bibinfo{pages}{022130} (\bibinfo{year}{2014}).

\bibitem[{\citenamefont{Sevilla and Sandoval}(2015)}]{SevillaPRE2015}
\bibinfo{author}{\bibfnamefont{F.~J.} \bibnamefont{Sevilla}} \bibnamefont{and}
  \bibinfo{author}{\bibfnamefont{M.}~\bibnamefont{Sandoval}},
  \bibinfo{journal}{Phys. Rev. E} \textbf{\bibinfo{volume}{91}},
  \bibinfo{pages}{052150} (\bibinfo{year}{2015}),
  \urlprefix\url{http://link.aps.org/doi/10.1103/PhysRevE.91.052150}.

\bibitem[{\citenamefont{Sevilla}(2016)}]{SevillaPRE2016}
\bibinfo{author}{\bibfnamefont{F.~J.} \bibnamefont{Sevilla}},
  \bibinfo{journal}{Phys. Rev. E} \textbf{\bibinfo{volume}{94}},
  \bibinfo{pages}{062120} (\bibinfo{year}{2016}),
  \urlprefix\url{http://link.aps.org/doi/10.1103/PhysRevE.94.062120}.

\bibitem[{\citenamefont{Mehraeen et~al.}(2008)\citenamefont{Mehraeen,
  Sudhanshu, Koslover, and Spakowitz}}]{MehraeenPRE2008}
\bibinfo{author}{\bibfnamefont{S.}~\bibnamefont{Mehraeen}},
  \bibinfo{author}{\bibfnamefont{B.}~\bibnamefont{Sudhanshu}},
  \bibinfo{author}{\bibfnamefont{E.~F.} \bibnamefont{Koslover}},
  \bibnamefont{and} \bibinfo{author}{\bibfnamefont{A.~J.}
  \bibnamefont{Spakowitz}}, \bibinfo{journal}{Phys. Rev. E}
  \textbf{\bibinfo{volume}{77}}, \bibinfo{pages}{061803}
  (\bibinfo{year}{2008}),
  \urlprefix\url{https://link.aps.org/doi/10.1103/PhysRevE.77.061803}.

\bibitem[{\citenamefont{Detcheverry}(2014)}]{DetcheverryEPJE2014}
\bibinfo{author}{\bibfnamefont{F.}~\bibnamefont{Detcheverry}},
  \bibinfo{journal}{The European Physical Journal E}
  \textbf{\bibinfo{volume}{37}}, \bibinfo{pages}{114} (\bibinfo{year}{2014}),
  ISSN \bibinfo{issn}{1292-895X},
  \urlprefix\url{https://doi.org/10.1140/epje/i2014-14114-2}.

\bibitem[{\citenamefont{Kenkre and Sevilla}(2007)}]{KenkreSevilla2007}
\bibinfo{author}{\bibfnamefont{V.}~\bibnamefont{Kenkre}} \bibnamefont{and}
  \bibinfo{author}{\bibfnamefont{F.~J.} \bibnamefont{Sevilla}}, in
  \emph{\bibinfo{booktitle}{{Contributions to Mathematical Physics: a Tribute
  to Gerard G. Emch TS. Ali, KB. Sinha, eds.}}} (\bibinfo{publisher}{{Hindustan
  Book Agency, New Delhi}}, \bibinfo{year}{2007}), pp.
  \bibinfo{pages}{147--160}.

\bibitem[{\citenamefont{Giuggioli et~al.}(2009)\citenamefont{Giuggioli,
  Sevilla, and Kenkre}}]{giuggioli2009generalized}
\bibinfo{author}{\bibfnamefont{L.}~\bibnamefont{Giuggioli}},
  \bibinfo{author}{\bibfnamefont{F.}~\bibnamefont{Sevilla}}, \bibnamefont{and}
  \bibinfo{author}{\bibfnamefont{V.}~\bibnamefont{Kenkre}},
  \bibinfo{journal}{Journal of Physics A: Mathematical and Theoretical}
  \textbf{\bibinfo{volume}{42}}, \bibinfo{pages}{434004}
  (\bibinfo{year}{2009}).

\bibitem[{\citenamefont{Goldstein}(1951)}]{GoldsteinQJMAM1951}
\bibinfo{author}{\bibfnamefont{S.}~\bibnamefont{Goldstein}},
  \bibinfo{journal}{The Quarterly Journal of Mechanics and Applied Mathematics}
  \textbf{\bibinfo{volume}{4}}, \bibinfo{pages}{129} (\bibinfo{year}{1951}),
  \eprint{http://qjmam.oxfordjournals.org/content/4/2/129.full.pdf+html},
  \urlprefix\url{http://qjmam.oxfordjournals.org/content/4/2/129.abstract}.

\bibitem[{\citenamefont{Zaburdaev et~al.}(2015)\citenamefont{Zaburdaev,
  Denisov, and Klafter}}]{ZaburdaevRMP2015}
\bibinfo{author}{\bibfnamefont{V.}~\bibnamefont{Zaburdaev}},
  \bibinfo{author}{\bibfnamefont{S.}~\bibnamefont{Denisov}}, \bibnamefont{and}
  \bibinfo{author}{\bibfnamefont{J.}~\bibnamefont{Klafter}},
  \bibinfo{journal}{Rev. Mod. Phys.} \textbf{\bibinfo{volume}{87}},
  \bibinfo{pages}{483} (\bibinfo{year}{2015}),
  \urlprefix\url{https://link.aps.org/doi/10.1103/RevModPhys.87.483}.

\bibitem[{\citenamefont{Larralde}(1997)}]{LarraldePRE1997}
\bibinfo{author}{\bibfnamefont{H.}~\bibnamefont{Larralde}},
  \bibinfo{journal}{Phys. Rev. E} \textbf{\bibinfo{volume}{56}},
  \bibinfo{pages}{5004} (\bibinfo{year}{1997}),
  \urlprefix\url{http://link.aps.org/doi/10.1103/PhysRevE.56.5004}.

\bibitem[{\citenamefont{Weber et~al.}(2011)\citenamefont{Weber, Radtke,
  Schimansky-Geier, and H{\"a}nggi}}]{WeberPRE2011}
\bibinfo{author}{\bibfnamefont{C.}~\bibnamefont{Weber}},
  \bibinfo{author}{\bibfnamefont{P.~K.} \bibnamefont{Radtke}},
  \bibinfo{author}{\bibfnamefont{L.}~\bibnamefont{Schimansky-Geier}},
  \bibnamefont{and}
  \bibinfo{author}{\bibfnamefont{P.}~\bibnamefont{H{\"a}nggi}},
  \bibinfo{journal}{Physical Review E} \textbf{\bibinfo{volume}{84}},
  \bibinfo{pages}{011132} (\bibinfo{year}{2011}).

\bibitem[{\citenamefont{Mardia}(1974)}]{Mardia74p115}
\bibinfo{author}{\bibfnamefont{K.~V.} \bibnamefont{Mardia}},
  \bibinfo{journal}{Sankhy{\=a}: The Indian Journal of Statistics, Series B}
  pp. \bibinfo{pages}{115--128} (\bibinfo{year}{1974}).

\bibitem[{\citenamefont{i~Wu et~al.}(2000)\citenamefont{i~Wu, Li, Springer, and
  Neill}}]{WuEcoMod2000}
\bibinfo{author}{\bibfnamefont{H.}~\bibnamefont{i~Wu}},
  \bibinfo{author}{\bibfnamefont{B.-L.} \bibnamefont{Li}},
  \bibinfo{author}{\bibfnamefont{T.~A.} \bibnamefont{Springer}},
  \bibnamefont{and} \bibinfo{author}{\bibfnamefont{W.~H.} \bibnamefont{Neill}},
  \bibinfo{journal}{Ecological Modelling} \textbf{\bibinfo{volume}{132}},
  \bibinfo{pages}{115 } (\bibinfo{year}{2000}), ISSN \bibinfo{issn}{0304-3800},
  \urlprefix\url{http://www.sciencedirect.com/science/article/pii/S0304380000003094}.

\bibitem[{\citenamefont{Schnitzer}(1993)}]{SchnitzerPRE1993}
\bibinfo{author}{\bibfnamefont{M.~J.} \bibnamefont{Schnitzer}},
  \bibinfo{journal}{Phys. Rev. E} \textbf{\bibinfo{volume}{48}},
  \bibinfo{pages}{2553} (\bibinfo{year}{1993}),
  \urlprefix\url{http://link.aps.org/doi/10.1103/PhysRevE.48.2553}.

\bibitem[{\citenamefont{Hancock and Baskaran}(2015)}]{HancockPRE2015}
\bibinfo{author}{\bibfnamefont{B.}~\bibnamefont{Hancock}} \bibnamefont{and}
  \bibinfo{author}{\bibfnamefont{A.}~\bibnamefont{Baskaran}},
  \bibinfo{journal}{Phys. Rev. E} \textbf{\bibinfo{volume}{92}},
  \bibinfo{pages}{052143} (\bibinfo{year}{2015}),
  \urlprefix\url{http://link.aps.org/doi/10.1103/PhysRevE.92.052143}.

\bibitem[{\citenamefont{Jones and Pewsey}(2005)}]{JonesCircleDistributions2005}
\bibinfo{author}{\bibfnamefont{M.~C.} \bibnamefont{Jones}} \bibnamefont{and}
  \bibinfo{author}{\bibfnamefont{A.}~\bibnamefont{Pewsey}},
  \bibinfo{journal}{Journal of the American Statistical Association}
  \textbf{\bibinfo{volume}{100}}, \bibinfo{pages}{1422} (\bibinfo{year}{2005}),
  \eprint{http://dx.doi.org/10.1198/016214505000000286},
  \urlprefix\url{http://dx.doi.org/10.1198/016214505000000286}.

\bibitem[{\citenamefont{Gro{\ss}mann et~al.}(2016)\citenamefont{Gro{\ss}mann,
  Peruani, and B\"ar}}]{GrossmannNJP2016}
\bibinfo{author}{\bibfnamefont{R.}~\bibnamefont{Gro{\ss}mann}},
  \bibinfo{author}{\bibfnamefont{F.}~\bibnamefont{Peruani}}, \bibnamefont{and}
  \bibinfo{author}{\bibfnamefont{M.}~\bibnamefont{B\"ar}},
  \bibinfo{journal}{New Journal of Physics} \textbf{\bibinfo{volume}{18}},
  \bibinfo{pages}{043009} (\bibinfo{year}{2016}),
  \urlprefix\url{https://doi.org/10.1088%2F1367-2630%2F18%2F4%2F043009}.

\bibitem[{\citenamefont{Crenshaw}(1993)}]{Crenshaw1BullMathBio1993}
\bibinfo{author}{\bibfnamefont{H.~C.} \bibnamefont{Crenshaw}},
  \bibinfo{journal}{Bulletin of Mathematical Biology}
  \textbf{\bibinfo{volume}{55}}, \bibinfo{pages}{197 } (\bibinfo{year}{1993}),
  ISSN \bibinfo{issn}{0092-8240},
  \urlprefix\url{http://www.sciencedirect.com/science/article/pii/S0092824005800692}.

\bibitem[{\citenamefont{Lauga et~al.}(2006)\citenamefont{Lauga, DiLuzio,
  Whitesides, and Stone}}]{LaugaBioPhysJ2006}
\bibinfo{author}{\bibfnamefont{E.}~\bibnamefont{Lauga}},
  \bibinfo{author}{\bibfnamefont{W.~R.} \bibnamefont{DiLuzio}},
  \bibinfo{author}{\bibfnamefont{G.~M.} \bibnamefont{Whitesides}},
  \bibnamefont{and} \bibinfo{author}{\bibfnamefont{H.~A.} \bibnamefont{Stone}},
  \bibinfo{journal}{Biophysical Journal} \textbf{\bibinfo{volume}{90}},
  \bibinfo{pages}{400} (\bibinfo{year}{2006}), ISSN \bibinfo{issn}{0006-3495},
  \urlprefix\url{http://www.sciencedirect.com/science/article/pii/S0006349506722214}.

\bibitem[{\citenamefont{Shenoy et~al.}(2007)\citenamefont{Shenoy, Tambe,
  Prasad, and Theriot}}]{ShenoyPNAS2007}
\bibinfo{author}{\bibfnamefont{V.~B.} \bibnamefont{Shenoy}},
  \bibinfo{author}{\bibfnamefont{D.~T.} \bibnamefont{Tambe}},
  \bibinfo{author}{\bibfnamefont{A.}~\bibnamefont{Prasad}}, \bibnamefont{and}
  \bibinfo{author}{\bibfnamefont{J.~A.} \bibnamefont{Theriot}},
  \bibinfo{journal}{Proceedings of the National Academy of Sciences}
  \textbf{\bibinfo{volume}{104}}, \bibinfo{pages}{8229} (\bibinfo{year}{2007}),
  \eprint{http://www.pnas.org/content/104/20/8229.full.pdf},
  \urlprefix\url{http://www.pnas.org/content/104/20/8229.abstract}.

\bibitem[{\citenamefont{Schmidt et~al.}(2008)\citenamefont{Schmidt, van~der
  Gucht, Biesheuvel, Weinkamer, Helfer, and Fery}}]{SchmidtEBioPhysJ2008}
\bibinfo{author}{\bibfnamefont{S.}~\bibnamefont{Schmidt}},
  \bibinfo{author}{\bibfnamefont{J.}~\bibnamefont{van~der Gucht}},
  \bibinfo{author}{\bibfnamefont{P.~M.} \bibnamefont{Biesheuvel}},
  \bibinfo{author}{\bibfnamefont{R.}~\bibnamefont{Weinkamer}},
  \bibinfo{author}{\bibfnamefont{E.}~\bibnamefont{Helfer}}, \bibnamefont{and}
  \bibinfo{author}{\bibfnamefont{A.}~\bibnamefont{Fery}},
  \bibinfo{journal}{European Biophysics Journal} \textbf{\bibinfo{volume}{37}},
  \bibinfo{pages}{1361} (\bibinfo{year}{2008}), ISSN \bibinfo{issn}{1432-1017},
  \urlprefix\url{http://dx.doi.org/10.1007/s00249-008-0340-x}.

\bibitem[{\citenamefont{Friedrich and J\"ulicher}(2008)}]{FriedrichNJP2008}
\bibinfo{author}{\bibfnamefont{B.~M.} \bibnamefont{Friedrich}}
  \bibnamefont{and}
  \bibinfo{author}{\bibfnamefont{F.}~\bibnamefont{J\"ulicher}},
  \bibinfo{journal}{New Journal of Physics} \textbf{\bibinfo{volume}{10}},
  \bibinfo{pages}{123025} (\bibinfo{year}{2008}),
  \urlprefix\url{http://stacks.iop.org/1367-2630/10/i=12/a=123025}.

\bibitem[{\citenamefont{Marine et~al.}(2013)\citenamefont{Marine, Wheat, Ault,
  and Posner}}]{MarinePRE2013}
\bibinfo{author}{\bibfnamefont{N.~A.} \bibnamefont{Marine}},
  \bibinfo{author}{\bibfnamefont{P.~M.} \bibnamefont{Wheat}},
  \bibinfo{author}{\bibfnamefont{J.}~\bibnamefont{Ault}}, \bibnamefont{and}
  \bibinfo{author}{\bibfnamefont{J.~D.} \bibnamefont{Posner}},
  \bibinfo{journal}{Phys. Rev. E} \textbf{\bibinfo{volume}{87}},
  \bibinfo{pages}{052305} (\bibinfo{year}{2013}),
  \urlprefix\url{http://link.aps.org/doi/10.1103/PhysRevE.87.052305}.

\bibitem[{\citenamefont{K\"ummel et~al.}(2013)\citenamefont{K\"ummel, ten
  Hagen, Wittkowski, Buttinoni, Eichhorn, Volpe, L\"owen, and
  Bechinger}}]{KummelPRL2013}
\bibinfo{author}{\bibfnamefont{F.}~\bibnamefont{K\"ummel}},
  \bibinfo{author}{\bibfnamefont{B.}~\bibnamefont{ten Hagen}},
  \bibinfo{author}{\bibfnamefont{R.}~\bibnamefont{Wittkowski}},
  \bibinfo{author}{\bibfnamefont{I.}~\bibnamefont{Buttinoni}},
  \bibinfo{author}{\bibfnamefont{R.}~\bibnamefont{Eichhorn}},
  \bibinfo{author}{\bibfnamefont{G.}~\bibnamefont{Volpe}},
  \bibinfo{author}{\bibfnamefont{H.}~\bibnamefont{L\"owen}}, \bibnamefont{and}
  \bibinfo{author}{\bibfnamefont{C.}~\bibnamefont{Bechinger}},
  \bibinfo{journal}{Phys. Rev. Lett.} \textbf{\bibinfo{volume}{110}},
  \bibinfo{pages}{198302} (\bibinfo{year}{2013}),
  \urlprefix\url{http://link.aps.org/doi/10.1103/PhysRevLett.110.198302}.

\bibitem[{\citenamefont{Takagi et~al.}(2013)\citenamefont{Takagi, Braunschweig,
  Zhang, and Shelley}}]{TakagiPRL2013}
\bibinfo{author}{\bibfnamefont{D.}~\bibnamefont{Takagi}},
  \bibinfo{author}{\bibfnamefont{A.~B.} \bibnamefont{Braunschweig}},
  \bibinfo{author}{\bibfnamefont{J.}~\bibnamefont{Zhang}}, \bibnamefont{and}
  \bibinfo{author}{\bibfnamefont{M.~J.} \bibnamefont{Shelley}},
  \bibinfo{journal}{Phys. Rev. Lett.} \textbf{\bibinfo{volume}{110}},
  \bibinfo{pages}{038301} (\bibinfo{year}{2013}),
  \urlprefix\url{https://link.aps.org/doi/10.1103/PhysRevLett.110.038301}.

\bibitem[{\citenamefont{Gomez-Solano et~al.}(2016)\citenamefont{Gomez-Solano,
  Blokhuis, and Bechinger}}]{Gomez-SolanoPRL2016}
\bibinfo{author}{\bibfnamefont{J.~R.} \bibnamefont{Gomez-Solano}},
  \bibinfo{author}{\bibfnamefont{A.}~\bibnamefont{Blokhuis}}, \bibnamefont{and}
  \bibinfo{author}{\bibfnamefont{C.}~\bibnamefont{Bechinger}},
  \bibinfo{journal}{Phys. Rev. Lett.} \textbf{\bibinfo{volume}{116}},
  \bibinfo{pages}{138301} (\bibinfo{year}{2016}),
  \urlprefix\url{https://link.aps.org/doi/10.1103/PhysRevLett.116.138301}.

\bibitem[{\citenamefont{Narinder et~al.}(2018)\citenamefont{Narinder,
  Bechinger, and Gomez-Solano}}]{NarinderPRL2018}
\bibinfo{author}{\bibfnamefont{N.}~\bibnamefont{Narinder}},
  \bibinfo{author}{\bibfnamefont{C.}~\bibnamefont{Bechinger}},
  \bibnamefont{and} \bibinfo{author}{\bibfnamefont{J.~R.}
  \bibnamefont{Gomez-Solano}}, \bibinfo{journal}{Phys. Rev. Lett.}
  \textbf{\bibinfo{volume}{121}}, \bibinfo{pages}{078003}
  (\bibinfo{year}{2018}),
  \urlprefix\url{https://link.aps.org/doi/10.1103/PhysRevLett.121.078003}.

\bibitem[{\citenamefont{van Teeffelen and L\"owen}(2008)}]{VanTeeffelenPRE2008}
\bibinfo{author}{\bibfnamefont{S.}~\bibnamefont{van Teeffelen}}
  \bibnamefont{and} \bibinfo{author}{\bibfnamefont{H.}~\bibnamefont{{L\"owen}}}, \bibinfo{journal}{Phys. Rev. E} \textbf{\bibinfo{volume}{78}},
  \bibinfo{pages}{020101(R)} (\bibinfo{year}{2008}),
  \urlprefix\url{http://link.aps.org/doi/10.1103/PhysRevE.78.020101}.

\bibitem[{\citenamefont{Friedrich and J\"ulicher}(2009)}]{FriedrichPRL2009}
\bibinfo{author}{\bibfnamefont{B.~M.} \bibnamefont{Friedrich}}
  \bibnamefont{and}
  \bibinfo{author}{\bibfnamefont{F.}~\bibnamefont{J\"ulicher}},
  \bibinfo{journal}{Phys. Rev. Lett.} \textbf{\bibinfo{volume}{103}},
  \bibinfo{pages}{068102} (\bibinfo{year}{2009}),
  \urlprefix\url{http://link.aps.org/doi/10.1103/PhysRevLett.103.068102}.

\bibitem[{\citenamefont{Ohta and Ohkuma}(2009)}]{OhtaPRL2009}
\bibinfo{author}{\bibfnamefont{T.}~\bibnamefont{Ohta}} \bibnamefont{and}
  \bibinfo{author}{\bibfnamefont{T.}~\bibnamefont{Ohkuma}},
  \bibinfo{journal}{Phys. Rev. Lett.} \textbf{\bibinfo{volume}{102}},
  \bibinfo{pages}{154101} (\bibinfo{year}{2009}),
  \urlprefix\url{https://link.aps.org/doi/10.1103/PhysRevLett.102.154101}.

\bibitem[{\citenamefont{Wittkowski and L\"owen}(2012)}]{WittkowskiPRE2012}
\bibinfo{author}{\bibfnamefont{R.}~\bibnamefont{Wittkowski}} \bibnamefont{and}
  \bibinfo{author}{\bibfnamefont{H.}~\bibnamefont{L\"owen}},
  \bibinfo{journal}{Phys. Rev. E} \textbf{\bibinfo{volume}{85}},
  \bibinfo{pages}{021406} (\bibinfo{year}{2012}),
  \urlprefix\url{http://link.aps.org/doi/10.1103/PhysRevE.85.021406}.

\bibitem[{\citenamefont{Ledesma-Aguilar
  et~al.}(2012)\citenamefont{Ledesma-Aguilar, L{\"o}wen, and
  Yeomans}}]{Ledesma-AguilarEPJE2012}
\bibinfo{author}{\bibfnamefont{R.}~\bibnamefont{Ledesma-Aguilar}},
  \bibinfo{author}{\bibfnamefont{H.}~\bibnamefont{L{\"o}wen}},
  \bibnamefont{and} \bibinfo{author}{\bibfnamefont{J.~M.}
  \bibnamefont{Yeomans}}, \bibinfo{journal}{The European Physical Journal E}
  \textbf{\bibinfo{volume}{35}}, \bibinfo{pages}{1} (\bibinfo{year}{2012}),
  ISSN \bibinfo{issn}{1292-895X},
  \urlprefix\url{http://dx.doi.org/10.1140/epje/i2012-12070-5}.

\bibitem[{\citenamefont{L{\"o}wen}(2016)}]{LowenEPJST2016}
\bibinfo{author}{\bibfnamefont{H.}~\bibnamefont{L{\"o}wen}},
  \bibinfo{journal}{The European Physical Journal Special Topics}
  \textbf{\bibinfo{volume}{225}}, \bibinfo{pages}{2319} (\bibinfo{year}{2016}),
  ISSN \bibinfo{issn}{1951-6401},
  \urlprefix\url{https://doi.org/10.1140/epjst/e2016-60054-6}.

\bibitem[{\citenamefont{Kurzthaler and
  Franosch}(2017)}]{KurzthalerSoftMatter2017}
\bibinfo{author}{\bibfnamefont{C.}~\bibnamefont{Kurzthaler}} \bibnamefont{and}
  \bibinfo{author}{\bibfnamefont{T.}~\bibnamefont{Franosch}},
  \bibinfo{journal}{Soft Matter} pp.~\bibinfo{pages}{--}
  (\bibinfo{year}{2017}), \urlprefix\url{http://dx.doi.org/10.1039/C7SM00873B}.

\bibitem[{\citenamefont{Duffy and Ford}(1997)}]{DuffyJBacter1997}
\bibinfo{author}{\bibfnamefont{K.~J.} \bibnamefont{Duffy}} \bibnamefont{and}
  \bibinfo{author}{\bibfnamefont{R.~M.} \bibnamefont{Ford}},
  \bibinfo{journal}{Journal of bacteriology} \textbf{\bibinfo{volume}{179}},
  \bibinfo{pages}{1428} (\bibinfo{year}{1997}).

  \bibitem[{\citenamefont{Wang et~al.}(2009)\citenamefont{Wang, Anthony, Bae, and
  Granick}}]{WangPNAS2009}
\bibinfo{author}{\bibfnamefont{B.}~\bibnamefont{Wang}},
  \bibinfo{author}{\bibfnamefont{S.~M.} \bibnamefont{Anthony}},
  \bibinfo{author}{\bibfnamefont{S.~C.} \bibnamefont{Bae}}, \bibnamefont{and}
  \bibinfo{author}{\bibfnamefont{S.}~\bibnamefont{Granick}},
  \bibinfo{journal}{Proceedings of the National Academy of Sciences}
  \textbf{\bibinfo{volume}{106}}, \bibinfo{pages}{15160}
  (\bibinfo{year}{2009}),
  \eprint{http://www.pnas.org/content/106/36/15160.full.pdf},
  \urlprefix\url{http://www.pnas.org/content/106/36/15160.abstract}.

\bibitem[{\citenamefont{{Wang Bo} et~al.}(2012)\citenamefont{{Wang Bo}, {Kuo
  James}, {Bae Sung Chul}, and {Granick Steve}}}]{WangNatureMat2012}
\bibinfo{author}{\bibnamefont{{Wang Bo}}}, \bibinfo{author}{\bibnamefont{{Kuo
  James}}}, \bibinfo{author}{\bibnamefont{{Bae Sung Chul}}}, \bibnamefont{and}
  \bibinfo{author}{\bibnamefont{{Granick Steve}}}, \bibinfo{journal}{Nat Mater}
  \textbf{\bibinfo{volume}{11}}, \bibinfo{pages}{481} (\bibinfo{year}{2012}),
  ISSN \bibinfo{issn}{1476-1122}, \bibinfo{note}{10.1038/nmat3308}.

\bibitem[{\citenamefont{Bhattacharya et~al.}(2013)\citenamefont{Bhattacharya,
  Sharma, Saurabh, De, Sain, Nandi, and Chowdhury}}]{BhattacharyaJPCB2013}
\bibinfo{author}{\bibfnamefont{S.}~\bibnamefont{Bhattacharya}},
  \bibinfo{author}{\bibfnamefont{D.~K.} \bibnamefont{Sharma}},
  \bibinfo{author}{\bibfnamefont{S.}~\bibnamefont{Saurabh}},
  \bibinfo{author}{\bibfnamefont{S.}~\bibnamefont{De}},
  \bibinfo{author}{\bibfnamefont{A.}~\bibnamefont{Sain}},
  \bibinfo{author}{\bibfnamefont{A.}~\bibnamefont{Nandi}}, \bibnamefont{and}
  \bibinfo{author}{\bibfnamefont{A.}~\bibnamefont{Chowdhury}},
  \bibinfo{journal}{The Journal of Physical Chemistry B}
  \textbf{\bibinfo{volume}{117}}, \bibinfo{pages}{7771} (\bibinfo{year}{2013}),
  \bibinfo{note}{pMID: 23777572}, \eprint{http://dx.doi.org/10.1021/jp401704e},
  \urlprefix\url{http://dx.doi.org/10.1021/jp401704e}.

\bibitem[{\citenamefont{Cressoni et~al.}(2012)\citenamefont{Cressoni,
  Viswanathan, Ferreira, and da~Silva}}]{CressoniPRE2012}
\bibinfo{author}{\bibfnamefont{J.~C.} \bibnamefont{Cressoni}},
  \bibinfo{author}{\bibfnamefont{G.~M.} \bibnamefont{Viswanathan}},
  \bibinfo{author}{\bibfnamefont{A.~S.} \bibnamefont{Ferreira}},
  \bibnamefont{and} \bibinfo{author}{\bibfnamefont{M.~A.~A.}
  \bibnamefont{da~Silva}}, \bibinfo{journal}{Phys. Rev. E}
  \textbf{\bibinfo{volume}{86}}, \bibinfo{pages}{022103}
  (\bibinfo{year}{2012}),
  \urlprefix\url{http://link.aps.org/doi/10.1103/PhysRevE.86.022103}.

\bibitem[{\citenamefont{Chubynsky and Slater}(2014)}]{ChubynskyPRL2014}
\bibinfo{author}{\bibfnamefont{M.~V.} \bibnamefont{Chubynsky}}
  \bibnamefont{and} \bibinfo{author}{\bibfnamefont{G.~W.}
  \bibnamefont{Slater}}, \bibinfo{journal}{Phys. Rev. Lett.}
  \textbf{\bibinfo{volume}{113}}, \bibinfo{pages}{098302}
  (\bibinfo{year}{2014}),
  \urlprefix\url{http://link.aps.org/doi/10.1103/PhysRevLett.113.098302}.

\bibitem[{\citenamefont{Wang et~al.}(2016)\citenamefont{Wang, Zhang, and
  Zhao}}]{WangPRE2016}
\bibinfo{author}{\bibfnamefont{J.}~\bibnamefont{Wang}},
  \bibinfo{author}{\bibfnamefont{Y.}~\bibnamefont{Zhang}}, \bibnamefont{and}
  \bibinfo{author}{\bibfnamefont{H.}~\bibnamefont{Zhao}},
  \bibinfo{journal}{Phys. Rev. E} \textbf{\bibinfo{volume}{93}},
  \bibinfo{pages}{032144} (\bibinfo{year}{2016}),
  \urlprefix\url{http://link.aps.org/doi/10.1103/PhysRevE.93.032144}.
  
  \bibitem[{\citenamefont{Martens et~al.}(2012)\citenamefont{Martens, Angelani,
  Di~Leonardo, and Bocquet}}]{MartensEPJE2012}
\bibinfo{author}{\bibfnamefont{K.}~\bibnamefont{Martens}},
  \bibinfo{author}{\bibfnamefont{L.}~\bibnamefont{Angelani}},
  \bibinfo{author}{\bibfnamefont{R.}~\bibnamefont{Di~Leonardo}},
  \bibnamefont{and} \bibinfo{author}{\bibfnamefont{L.}~\bibnamefont{Bocquet}},
  \bibinfo{journal}{The European Physical Journal E}
  \textbf{\bibinfo{volume}{35}}, \bibinfo{pages}{84} (\bibinfo{year}{2012}),
  ISSN \bibinfo{issn}{1292-895X},
  \urlprefix\url{https://doi.org/10.1140/epje/i2012-12084-y}.

\bibitem[{\citenamefont{{Kurzthaler Christina}
  et~al.}(2016)\citenamefont{{Kurzthaler Christina}, {Leitmann Sebastian}, and
  {Franosch Thomas}}}]{KurzthalerSciRep2016}
\bibinfo{author}{\bibnamefont{{Kurzthaler Christina}}},
  \bibinfo{author}{\bibnamefont{{Leitmann Sebastian}}}, \bibnamefont{and}
  \bibinfo{author}{\bibnamefont{{Franosch Thomas}}},
  \bibinfo{journal}{Scientific Reports} \textbf{\bibinfo{volume}{6}},
  \bibinfo{pages}{36702} (\bibinfo{year}{2016}).

\end{thebibliography}
% \bibliographystyle{apsrev}

\end{document}